\documentclass[conference]{IEEEtran}

\usepackage{color}
\usepackage{algorithm,algorithmicx,algcompatible}
\usepackage{verbatim, subfigure}
\usepackage{array, calc, multicol}
\usepackage{xspace}
\usepackage{graphics, epstopdf, graphicx, epsfig, psfrag, epsf, subfigure}
\usepackage{amssymb, amsfonts, amsmath, times}
\usepackage{multicol,balance, verbatim}
\usepackage{textcomp}

\newcommand{\ie}{{\em i.e., }}

\newcommand{\eg}{{\em e.g., }}

\newtheorem{theorem}{Theorem}

\newtheorem{example}{Example}

\DeclareGraphicsExtensions{.eps, .pdf, .png, .jpg}   

\newcommand{\Sset}{\mathcal{S}}

\newcommand{\Nset}{\mathcal{N}}

\newcommand{\Lset}{\mathcal{L}}

\newcommand{\Uset}{\mathcal{U}}

\DeclareMathOperator*{\argmax}{arg\,max}

\IEEEoverridecommandlockouts

\makeatletter
\newcommand{\oset}[2]{%
{\mathop{#2}\limits^{\vbox to -.5\ex@{\kern-\tw@\ex@
\hbox{\scriptsize #1}\vss}}}}
\makeatother

\begin{document}

\title{TCP-Aware Backpressure Routing and Scheduling
\thanks{This work was supported by NSF grant CNS-0915988, ONR grant N00014-12-1-0064, ARO Muri grant number W911NF-08-1-0238.}
}


\author{Hulya Seferoglu and Eytan Modiano\\
{Laboratory For Information and Decision Systems}\\
{Massachusetts Institute of Technology}\\
{ \small \tt \{hseferog, modiano\}@mit.edu}\\
}

\maketitle

\begin{abstract}
In this work, we explore the performance of backpressure routing and scheduling for TCP flows over wireless networks. 
TCP and backpressure are not compatible 
due to a mismatch between the congestion control mechanism of TCP and the queue size based routing and scheduling of the backpressure framework. We propose a TCP-aware backpressure routing and scheduling that takes into account the behavior of TCP flows. TCP-aware backpressure (i) provides throughput optimality guarantees in the Lyapunov optimization framework, (ii) gracefully combines TCP and backpressure without making any changes to the TCP protocol, (iii) improves the throughput of TCP flows significantly, and (iv) provides fairness across competing TCP flows. 
\end{abstract}



\section{Introduction}\label{sec:intro}
The backpressure routing and scheduling paradigm has emerged from the pioneering work \cite{tass_eph1}, \cite{tass_eph2}, which showed that, in wireless networks where nodes route and schedule packets based on queue backlog differences, one can stabilize the queues for any feasible traffic. This seminal idea has generated a lot of research interest. Moreover, it has been shown that backpressure can be combined with flow control to provide utility-optimal operation \cite{neely_mod}.

The strengths of these techniques have recently increased the interest in practical implementation of the backpressure framework over wireless networks as summarized in Section~\ref{sec:related}. One important practical problem that remains open, and is the focus of this paper, is the performance of backpressure with Transmission Control Protocol (TCP) flows.


TCP is the dominant transport protocol in the Internet today and is likely to remain so for the foreseeable future. Therefore, it is crucial to exploit throughput improvement potential of backpressure routing and scheduling for TCP flows. However, TCP flows are not compatible with backpressure.
Their joint behavior is so detrimental that some flows may never get a chance to transmit. To better illustrate this point, we first discuss the operation of backpressure in the following example.

\begin{figure}[t!]
\vspace{-10pt}
\centering
\subfigure[\scriptsize Backpressure with random arrivals with rates $A_1(t)$, $A_2(t)$]{ \label{fig:intro_example_a} \scalebox{.65}{\includegraphics[bb=0 0 165 180]{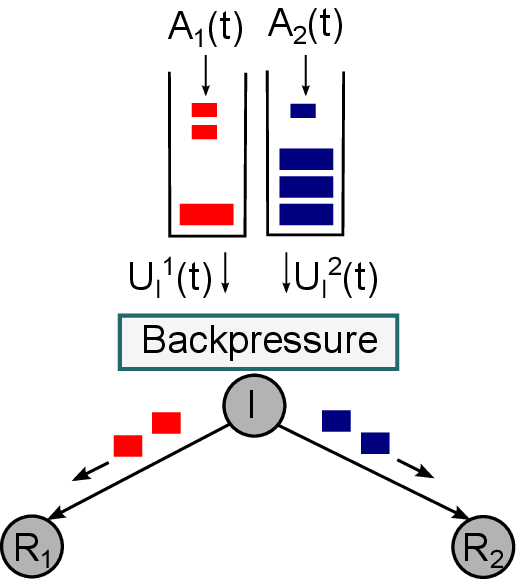}} }
\subfigure[\scriptsize Backpressure with TCP arrivals]{ \scalebox{.65}{\label{fig:intro_example_b}\includegraphics[bb=0 0 165 180]{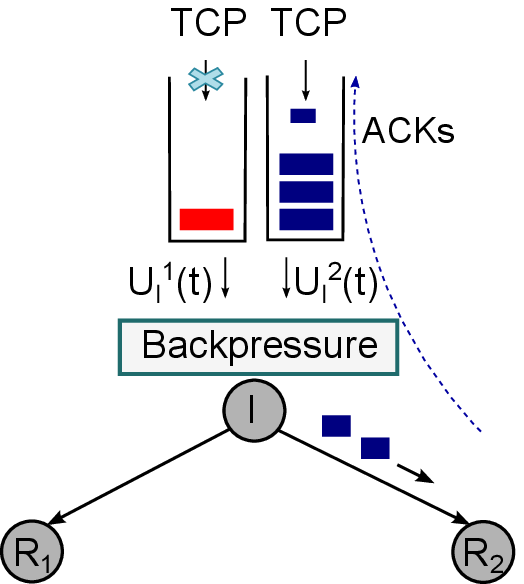}} }
\vspace{-5pt}
\caption{\scriptsize  Example one-hop downlink topology consisting of a transmitter node $I$, and two receiver nodes; $R_1$ and $R_2$. The two flows; $1$ and $2$ are destined to $R_1$ and $R_2$, respectively. $U_{I}^{1}(t)$ and $U_{I}^{2}(t)$ are per-flow queue sizes at time $t$. (a) Backpressure with random arrivals with rates $A_1(t)$ and $A_2(t)$. (b) Backpressure with TCP arrivals.  }
\vspace{-20pt}
\label{fig:intro_example}
\end{figure}

\begin{example}\label{ex1}
Let us consider Fig.~\ref{fig:intro_example}, which shows an example one-hop downlink topology consisting of a transmitter node $I$, and two receiver nodes; $R_1$ and $R_2$. The two flows; $1$ and $2$ are destined to $R_1$ and $R_2$, respectively. $U_{I}^{1}(t)$ and $U_{I}^{2}(t)$ are per-flow queue sizes at time $t$. Let us focus on Fig.~\ref{fig:intro_example}(a). At time $t$, packets from the two flows arrive according to random arrival rates; $A_1(t)$ and $A_2(t)$, respectively. The packets are stored in per-flow queues.
The backpressure scheduling algorithm, also known as max-weight scheduling, determines the queue (hence the flow) from which packets should be transmitted at time $t$. The decision is based on queue backlog differences, \ie $U_{I}^{1}(t) - U_{R_1}^{1}(t)$ and $U_{I}^{2}(t) - U_{R_2}^{2}(t)$, where $U_{R_1}^{1}(t)$ and $U_{R_2}^{2}(t)$ are per-flow queue sizes at $R_1$ and $R_2$, respectively. Since $R_1$ and $R_2$ are the destination nodes, the received packets are immediately passed to the higher layers, so $U_{R_1}^{1}(t) = U_{R_2}^{2}(t) = 0$, $\forall t$. Therefore, the scheduling algorithm makes the scheduling decision based on $U_{I}^{1}(t)$ and $U_{I}^{2}(t)$. In particular, the scheduling decision is $s^{*} = \argmax \{U_{I}^{1}(t),U_{I}^{2}(t)\}$ such that $s^{*} \in \{1,2\}$. Note that a packet(s) from flow $s^{*}$ is transmitted at time $t$. It was shown in \cite{tass_eph1}, \cite{tass_eph2} that if the arrivals rates $A_1(t)$ and $A_2(t)$ are inside the stability region, the scheduling algorithm stabilizes the queues. Note that the arrival rates $A_1(t)$ and $A_2(t)$ are independent from the scheduling decisions, \ie the scheduling decisions do not affect $A_1(t)$ and $A_2(t)$. However, this is not true if the flows are regulated by TCP as explained next.\hfill $\blacksquare$
\end{example}

The fundamental goal of TCP, which applies to all TCP variants, is to achieve as much bandwidth as possible while maintaining some level of long-term rate fairness across competing flows. By their very design, all TCP algorithms (both the widely employed loss-based versions and the delay-based ones) have their own ``clock'', which relies on end-to-end acknowledgement (ACK) packets. Based on the received ACKs, TCP determines whether and how many packets should be injected into the network by updating its window size.

\begin{figure}
\vspace{-10pt}
\centering
\scalebox{.4}{\includegraphics[bb=0 0 556 193]{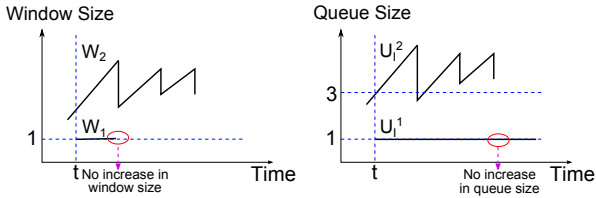}}
\vspace{-5pt}
\caption{\scriptsize Sample paths that show the evolution of $W_1$,$W_2$ and $U_{I}^{1}$, $U_{I}^{2}$ over time. Note that $W_1$,$W_2$ are the congestion window size of the TCP flows, and  $U_{I}^{1}$, $U_{I}^{2}$ are the corresponding queue sizes for the example presented in Fig.~\ref{fig:intro_example}(b). Due to backpressure, $W_1$ does not increase and $U_{I}^{1}$ does not receive or transmit any packets, and its size stays the same; $U_{I}^{1}(t) = 1, \forall t$.}
\label{fig:window_queue_evolution}
\vspace{-20pt}
\end{figure}

{\em Example 1 - continued:} Let us consider Fig.~\ref{fig:intro_example}(b) to illustrate the interaction of backpressure and TCP. In Fig.~\ref{fig:intro_example}(b), packet arrivals are controlled by TCP. Let us consider that loss-based TCP flavor, \eg TCP-Reno or TCP-SACK, is employed. Assume that at time $t$, the TCP congestion window size of the first flow, \ie $W_1(t)$, is small, \eg $W_1(t) = 1$ segment, (note that 1-segment window size may be seen at the beginning of a connection or after a re-transmit timeout), while the TCP congestion window size of the second flow is $W_2(t) > 1$ (\eg it may be the case that flow 2 has been transmitting for some time until $t$, and it has already increased its window size). As depicted in the figure, the example queue occupancies at time $t$ are $U_{I}^{1}(t) = 1$ and $U_{I}^{2}(t) = 3$. Since, $U_{I}^{2}(t) > U_{I}^{1}(t)$, a packet(s) from the second flow is transmitted. $R_2$ receives the transmitted packet, and passes it to TCP. TCP generates an ACK, and transmits it back to node $I$. TCP source of flow $2$ at node $I$ increases window size after receiving an ACK. Therefore, more packets are passed to $U_{I}^{2}(t)$. On the other hand, since $U_{I}^{1}(t) < U_{I}^{2}(t)$, no packets are transmitted from flow $1$. Thus, TCP does not receive any ACKs for the $1$st flow, does not increase its window size, and no (or sporadic) new packets are passed to $U_{I}^{1}(t)$. Eventually, the size of $U_{I}^{1}(t)$ almost never increases, so no packets are transmitted from flow $1$. Possible sample paths showing the evolution of $W_1$ and $W_2$ as well as $U_{I}^{1}$ and $U_{I}^{2}$ over time are shown in Fig.~\ref{fig:window_queue_evolution}. As can be seen, the joint behavior of TCP and backpressure is so detrimental that flow $1$ does not get any chance to transmit. We confirm this observation via simulations in Section~\ref{sec:performance}.\hfill $\blacksquare$

The incompatibility of backpressure is not limited to the loss-based versions of TCP. The delay-based TCP flavors, \eg TCP Vegas is also incompatible with backpressure, as TCP-Vegas has its own clock, which relies on end-to-end ACK packets to calculate round-trip-times (RTTs). If some packets are trapped in buffers due to backpressure as in the above example, sporadic or no ACK packets are received. This increases RTTs, and reduces end-to-end rate of TCP Vegas as there is inverse relationship between RTT and rate. Furthermore, backpressure leads to timeouts which reduce the end-to-end rate in both loss-based and delay-based TCP versions, including new TCP versions; TCP-Compound \cite{tcp_compound} and TCP-Cubic \cite{tcp_cubic}.

In this paper, we propose ``TCP-aware backpressure'' that helps TCP and backpressure operate in harmony. In particular, TCP-aware backpressure takes into account the behavior of TCP flows, and gives transmission opportunity to flows with short queues. This makes all TCP flows transmit their packets, so the TCP clock, which relies on packet transmissions and end-to-end ACKs, continues to operate. 
Furthermore, the throughput of TCP flows improves by exploiting the performance of the backpressure routing and scheduling. We note that backpressure introduces additional challenges when combined with TCP such as out of order delivery, high jitter RTTs, and packet losses due to corruption over wireless links. However, these challenges are not specific to backpressure, and exist when a multiple path routing scheme over wireless networks is combined with TCP. We address these challenges by employing network coding (in Section~\ref{sec:algs}). Yet, the main focus of this paper is the incompatibility of TCP and backpressure and developing a TCP-aware backpressure framework. 
The following are the key contributions of this work:

\begin{itemize}
  \item We identify the mismatch between TCP and the backpressure framework; \ie their joint behavior is so detrimental that some flows may never get a chance to transmit. In order to address the mismatch between TCP and backpressure, we develop ``TCP-aware backpressure routing and scheduling''.
  \item We show that (i) TCP-aware backpressure routing and scheduling stabilizes queues for any feasible traffic as the classical backpressure \cite{tass_eph1}, \cite{tass_eph2}, (ii) TCP-aware backpressure routing and scheduling provides the same utility-optimal operation guarantee when combined with a flow control algorithm as the classical backpressure \cite{neely_mod}.
  \item We provide implementation details and explain how to tune TCP-aware backpressure in practice so that it complies with TCP. Moreover, we combine network coding and TCP-aware backpressure to address the additional challenges such as out of order delivery, packet loss, and jitter. Thanks to employing network coding, which makes TCP flows sequence agnostic (with respect to packet IDs), TCP-aware backpressure fully complies with TCP.
  \item We evaluate our schemes in a multi-hop setting, using ns-2 \cite{ns2}. The simulation results (i) confirm the mismatch of TCP and backpressure, (ii) show that TCP-aware backpressure is compatible with TCP, and significantly improves throughput as compared to existing adaptive routing schemes, (iii) demonstrate that TCP-aware backpressure provides fairness across competing TCP flows. 
\end{itemize}

The structure of the rest of the paper is as follows. Section~\ref{sec:system} gives an overview of the system model. Section~\ref{sec:opt} presents TCP-aware backpressure design and analysis. Section~\ref{sec:algs} presents the implementation details of TCP-aware backpressure as well as its interaction with TCP. Section~\ref{sec:performance} presents simulation results.
Section~\ref{sec:related} presents related work. Section~\ref{sec:conclusion} concludes the paper.



\section{System Model}\label{sec:system}
\begin{figure}
\vspace{-10pt}
\centering
\scalebox{.58}{\includegraphics[bb=0 0 400 165]{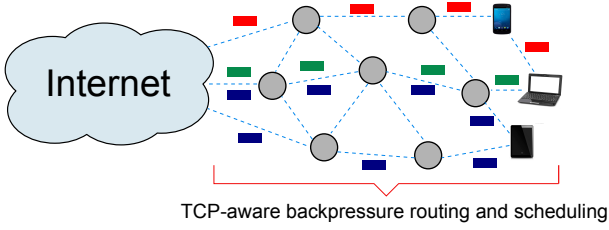}}
\vspace{-5pt}
\caption{\scriptsize A general network model that we consider in this paper. A flow may originate from a source in the Internet and traverse multiple hops to reach its destination in a wireless network. An end-to-end TCP connection is set up for each flow. We explore the performance of backpressure for TCP flows in the wireless network.}
\label{fig:main-example}
\vspace{-20pt}
\end{figure}

We consider a general network model presented in Fig.~\ref{fig:main-example}, where flows may originate from a source in the Internet and traverse multiple hops to reach their destination in a wireless network. An end-to-end TCP connection is set up for each flow.
Our goal in this paper is to develop TCP-aware backpressure routing and scheduling algorithms that operate in the wireless network. In this direction, we first develop our algorithms using the Lyapunov optimization framework (which is presented in Section~\ref{sec:opt}) by taking into account the incompatibility of TCP and classical backpressure. In this section, we provide an overview of the system model and assumptions that we use to develop the TCP-aware backpressure. Note that the interaction and implementation of TCP-aware backpressure routing and scheduling with actual TCP flows are presented in Section~\ref{sec:algs}.

{\em Wireless Network Setup:} The wireless network consists of $N$ nodes and $L$ links, where $\Nset$ is the set of nodes and $\Lset$ is the set of links in the network. In this setup, each wireless node is able to perform routing and scheduling. Let $\Sset$ be the set of unicast flows between source-destination pairs in the network.
We consider in our formulation and analysis that time is slotted, and $t$ refers to the beginning of slot $t$.

{\em Channel Model:} At slot $t$, $\boldsymbol C(t)$ $= \{C_{1}(t),$ $...,$ $C_{l}(t),$ $..., C_{L}(t)\}$ is the channel state vector, where $l$ represents the edges such that $l = (i,j)$, $(i,j) \in \Lset$ and $i \neq j$. For the sake of analysis, we assume that $C_{l}(t)$ is the state of link $l$ at time $t$ and takes values from the set $\{ON,OFF\}$ according to a probability distribution which is i.i.d. over time slots. If  $C_{l}(t) = ON$, packets can be transmitted with rate $R_l$. Otherwise; (\ie if $C_{l}(t) = OFF$), no packets are transmitted. Note that our analysis can be extended to more general channel state models \cite{neely_book}. We also consider a Rayleigh fading model in our simulations.

Let $\Gamma_{\boldsymbol C(t)}$ denote the set of the link transmission rates feasible at time slot $t$ and for channel state $\boldsymbol C(t)$ and interference among wireless links. In particular, at every time slot $t$, the link transmission vector $\boldsymbol f(t) = \{f_1(t), ..., f_l(t), ... f_L(t)\}$ should be constrained such that $\boldsymbol f(t)$ $\in \Gamma_{\boldsymbol C(t)}$. Hence, $f_l(t)$ takes a value from the set $\{R_l,0\}$ depending on the channel state and interference among multiple wireless nodes. Also note that $\boldsymbol f(t)$ is determined by the scheduling algorithm.

{\em Stability Region:}
Let $(\lambda_s)$ be the vector of arrival rates $\forall s \in \Sset$. The network stability region $\Lambda$ is defined as the closure of all arrival rate vectors that can be stably transmitted in the network, considering all possible routing and scheduling policies \cite{tass_eph1}, \cite{tass_eph2}, \cite{neely_mod}. $\Lambda$ is fixed and depends only on channel statistics and interference.


{\em Flow Rates and Queue Evolution:}  Each flow $s \in \Sset$ is generated at its source node according to an arrival process $A_s(t)$, $\forall s \in \Sset$ at time slot $t$. The arrivals are i.i.d. over the slots and $\lambda_s = E[A_s(t)]$, $\forall s \in \Sset$. We assume that $E[A_s(t)]$ and $E[A_s(t)^{2}]$ are finite. Note that we make i.i.d. arrivals assumption for the purpose of designing and analyzing our algorithms in the Lyapunov optimization framework. This assumption is relaxed in the practical setup when we combine our algorithms with TCP flows in Section~\ref{sec:algs}.

Each node $i$ constructs a per-flow queue $\Uset_{i}^{s}$ for each flow $s \in \Sset$. The size of the per-flow queue $\Uset_{i}^{s}$ at time $t$ is $U_{i}^{s}(t)$.
Let $o(s)$ be the source node of flow $s$. The packets generated according to the arrival process $A_s(t)$ are inserted in the per-flow queue at node $o(s)$, \ie in $\Uset_{o(s)}^{s}$. These queues only store packets from flow $s \in \Sset$. Each node $i$ such that $i \in \Nset$ and $i \neq o(s)$, may receive packets from its neighboring nodes and insert them in $\Uset_{i}^{s}$. The transmission rate of flow $s$ from node $i$ to node $j$ is $f_{i,j}^{s}(t)$. Since the link transmission rate over link $(i,j)$ is $f_{i,j}(t)$ at time $t$, multiple flows could share the available rate, \ie $\sum_{s \in \Sset} f_{i,j}^{s}(t) \leq f_{i,j}(t)$. Accordingly, at every time slot $t$, the size of per-flow queues, \ie $U_{i}^{s}(t)$ evolves according to the following dynamics.
\begin{align} \label{eq:queue_U}
& U_{i}^{s}(t+1) \leq \max [U_{i}^{s}(t) - \sum_{j \in \Nset} f_{i,j}^{s}(t), 0] + \sum_{j \in \Nset} f_{j,i}^{s}(t) \nonumber \\
& + A_{s}(t)1_{[i=o(s)]},
\end{align} where $1_{[i=o(s)]}$ is an indicator function, which is $1$ if $i=o(s)$, and $0$, otherwise. Note that Eq.~(\ref{eq:queue_U}) is inequality, because the number of packets in the queue $U_{j}^{s}(t)$ may be less than $ f_{j,i}^{s}(t)$.


\section{TCP-Aware Backpressure: Design and Analysis}\label{sec:opt}
In this section, we design and analyze the TCP-aware backpressure scheme. In particular, we provide a stochastic control strategy including routing and scheduling to address the incompatibility between TCP and classical backpressure.

\textbf{\underline{TCP-Aware Backpressure:}}
\begin{itemize}
  \item \textbf{Routing \& Intra-Node Scheduling.} The routing \& intra-node scheduling part of TCP-aware backpressure determines a flow $s$ from which packets should be transmitted at slot $t$ from node $i$, as well as the next hop node $j$ to which packets from flow $s$ should be forwarded. The algorithm works as follows.

  Node $i$ observes per-flow queue backlogs in all neighboring nodes at time $t$, and determines queue backlog difference according to:
  \begin{align} \label{eq:per_flow_difference}
  D_{i,j}^{s}(t) = \max\{K,U_{i}^{s}(t)\} - U_{j}^{s}(t),
  \end{align} where $K$ is a non-negative finite constant. Let $l = (i,j)$ s.t. $j \in \Nset$ and $j \neq i$. The maximum queue backlog difference among all flows over link $l \in \Lset$ is;
  \begin{align} \label{eq:per_link_difference}
  D_{l}^{*}(t) = \max_{[s \in \Sset | l \in \Lset_{s}]} \{ D_{l}^{s}(t) \}.
  \end{align}
  The flow that maximizes the queue backlog differences over link $l$ is $s_{l}^{*}(t)$ and expressed as;
  \begin{align} \label{eq:selected_flow}
  s_{l}^{*}(t) = \argmax_{[s \in \Sset | l \in \Lset_{s}]} \{ D_{l}^{s}(t) \}.
  \end{align}
  At time slot $t$, one or more packets are selected from the queue $\Uset_{i}^{s_{l}^{*}(t)}$ if $D_{l}^{*}(t)$ $>$ $0$ and $\Uset_{i}^{s_{l}^{*}(t)}$ has enough packets for transmission. The transmission of the selected packets depends on the channel conditions and interference constraints, and determined by inter-node scheduling.

  Note that TCP-aware backpressure uses queue backlog difference $\max\{K,U_{i}^{s}(t)\} - U_{j}^{s}(t)$ in Eq.~(\ref{eq:per_flow_difference}) instead of $U_{i}^{s}(t) - U_{j}^{s}(t)$ in the classical backpressure. The advantage of using Eq.~(\ref{eq:per_flow_difference}) in TCP-aware backpressure is that node $i$ may select packets from flow $s$ even if queue size $U_{i}^{s}(t)$ is small.\footnote{\scriptsize Note that place-holder backlogs, such as using $U_{i}^{s}(t)+K$ instead of $U_{i}^{s}(t)$ has been considered in the literature \cite{neely_book}. Although place-holder algorithms are beneficial to improve end-to-end delay, they do not solve the problem we consider in this paper as they do not give transmission opportunity to small queues.} This advantage is clarified through an illustrative example later in this section.
  \item \textbf{Inter-Node Scheduling.} The inter-node scheduling (as also called resource allocation \cite{neely_mod}) part of TCP-aware backpressure determines link transmission rates considering the link state information and interference constraints.

  Each node $i$ observes the channel state $\boldsymbol C(t)$ at time $t$, and determines a transmission vector $\boldsymbol f(t) = \{f_1(t), ...,$ $f_l(t), ... f_L(t)\}$ by maximizing $\sum_{l \in \Lset} D_{l}^{*}(t) f_l(t)$. Note that $\boldsymbol f(t)$ should be constrained such that $\boldsymbol f(t) \in \Gamma_{\boldsymbol C(t)}$, \ie interference among multiple nodes should be taken into account. The resulting transmission rate $f_{l}(t)$ is used to transmit packets of flow $s_{l}^{*}(t)$ over link $l$.
\end{itemize}
\begin{theorem}\label{theorem1}
If channel states are i.i.d. over time slots, the arrival rates $\lambda_s$, $\forall s \in \Sset$ are interior to the stability region $\Lambda$, and $K$ is a non-negative finite constant, then TCP-aware backpressure stabilizes the network and the total average queue size is bounded.
\end{theorem}
{\em Proof:} The proof is provided in Appendix A. \hfill $\blacksquare$
\begin{example}\label{ex2}
Let us consider again Fig.~\ref{fig:intro_example}(b) for the operation of TCP-aware backpressure. The example queue occupancies at time $t$ are $U_{I}^{1}(t) = 1$ and $U_{I}^{2}(t) = 3$. Assume that $K$ in Eq.~(\ref{eq:per_flow_difference}) is chosen as $K=10$. According to TCP-aware backpressure, the scheduling algorithm makes a decision based on the rule $s^{*} = \argmax \{\max \{K,U_{I}^{1}(t)\},\max$ $\{K,U_{I}^{2}(t)\}\}$ such that $s^{*} \in \{1,2\}$. Since $\max$ $\{K,U_{I}^{s}(t)\}$ $=$ $10$, $s=1,2$, both flows get equal chance for transmission. Thus, congestion window sizes of both TCP flows evolve in time, and the TCP flows can transmit their packets. We note that one can extend this example for the case; $U_{I}^{1}(t) = 7$ and $U_{I}^{2}(t) = 12$. In this case, as $K=10$, packets from the first flow may not get any chance for transmission. Therefore, it is crucial to determine $K$ in practice, which we explain in Section~\ref{sec:algs}. \hfill $\blacksquare$
\end{example}

Note that we propose TCP-aware backpressure; its routing, intra-node scheduling, and inter-node scheduling parts to work with TCP and TCP's end-to-end flow control mechanism. In the next section, we provide implementation details. However, TCP-aware backpressure can also be combined with flow control schemes other than TCP's, which is important for two reasons: (i) it may be possible or preferable to use personalized flow control mechanisms instead of TCP's in some systems, (ii) there may be both TCP and non-TCP flows in some systems, where a TCP-friendly flow control mechanism combined with non-TCP flows is crucial to accommodate both TCP and non-TCP flows. We consider the following flow control algorithm, developed in \cite{neely_mod}, to complement TCP-aware backpressure for non-TCP flows.

The flow control algorithm at node $i$ determines the number of packets from flow $s$ that should be passed to the per-flow queues; $\Uset_{i}^{s}$ at every time slot $t$ according to;

\begin{align} \label{eq:flow_control}
\max_{\boldsymbol x} & \sum_{[s \in \Sset | i=o(s)]} [Mg_{s}(x_s(t)) -  U_{i}^{s}(t) x_{s}(t) ] \nonumber \\
\mbox{s.t. } &  \sum_{[s \in \Sset  | i=o(s)]} x_{s}(t) \leq R_{i}^{max}
\end{align} where $R_{i}^{max}$ is a constant larger than the maximum outgoing rate from node $i$, $M$ is a positive constant, $x_s(t)$ is the rate of packets that will be inserted to the per-flow queue $\Uset_{i}^{s}$, and $g_{s}(.)$ is the utility function of flow $s$.

\begin{theorem}\label{theorem2}
If there are only non-TCP flows in the system and they employ the flow control algorithm in Eq.~(\ref{eq:flow_control}) and TCP-aware backpressure (with non-negative finite value of $K$ in Eq.~(\ref{eq:per_flow_difference})), then the admitted flow rates converges to the utility optimal operating point (as the classical backpressure) in the stability region $\Lambda$ with increasing $M$.
\end{theorem}
{\em Proof:} 
The proof of Theorem~\ref{theorem2} directly follows when Appendix A and drif+penalty approach \cite{neely_mod} are combined. \hfill $\blacksquare$

\section{TCP-Aware Backpressure: Implementation \& Interaction with TCP}\label{sec:algs}
\begin{figure}
\hspace{-5pt}
\centering
\scalebox{.38}{\includegraphics[bb=0 0 576 272]{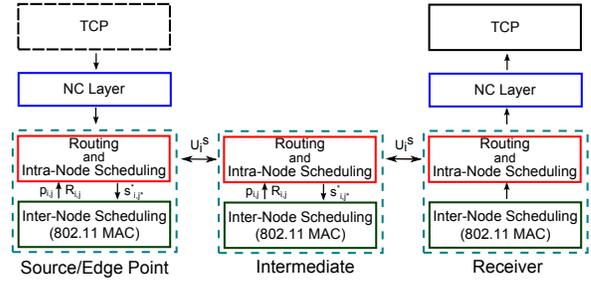}}
\caption{\scriptsize  TCP-aware backpressure operations at edge-points and intermediate nodes. The {\em inter-node scheduling} and {\em routing and intra-node scheduling} parts of TCP are implemented on top of 802.11 MAC and in network layers, respectively. The {\em NC layer} is implemented as a slim layer above the network layer at the edge points. Transport layer, \ie TCP, only exists if the edge point is a TCP source.
}
\vspace{-10pt}
\label{fig:protocol_stack}
\vspace{-20pt}
\end{figure}

We present practical implementation details of TCP-aware backpressure as well as its interaction with different layers in the protocol stack (summarized in Fig.~\ref{fig:protocol_stack}).

\subsection{Implementation}
\subsubsection{Inter-Node Scheduling}
The inter-node scheduling part of TCP-aware backpressure determines which links should be activated at time $t$.  The inter-node scheduling is a hard problem, \cite{tutorial_doyle}, \cite{lin_schroff_tutorial}, so its practical implementation is challenging. Therefore, we implement its low complexity version in our system on top of IEEE 802.11 MAC as seen in Fig.~\ref{fig:protocol_stack}. The implementation details are as follows.




Each node uses 802.11 MAC to access the wireless medium. When a node $i$ is assigned a channel by the MAC protocol, inter-node scheduling determines the neighboring node to which a selected packet should be forwarded. Let us assume that a packet is selected from flow $s_{i,j}^{*}(t)$ to be forwarded to node $j$ by the routing and intra-node scheduling algorithm, which we explain later in this section. The next hop that the selected packet should be forwarded is $j^{*}$ and determined by $j^{*} = \argmax_{j \in \Nset} \{D_{i,j}^{*} \tilde{R}_{i,j} (1 - \tilde{p}_{i,j})\}$, where $\tilde{p}_l$ and $\tilde{R}_l$ are the estimated values of $p_l$ (loss probability) and $R_l$ (link transmission rate) over link $l=(i,j)$, respectively.\footnote{\scriptsize $\tilde{p}_l$ is calculated as one minus the ratio of successfully transmitted packets over all transmitted packets during a time interval $T$ on link $l$. $\tilde{R}_l$ is calculated as the average of the recent (over an interval) link rates over link $l$.}
Then, a packet from flow $s_{i,j^{*}}^{*}(t)$, \ie from the network layer queue $U_{i}^{s_{i,j^{*}}^{*}(t)}$, is removed and passed to the MAC layer for transmission. The MAC layer transmits the packet to node $j^{*}$.


\subsubsection{Routing and Intra-Node Scheduling}
This algorithm determines the next hop(s) to which packets should be forwarded, and the packets that should be transmitted.

We construct per-flow queues, \ie $\Uset_{i}^{s}$, at the network layer\footnote{\scriptsize Note that constructing per-flow queues at each node may not be feasible in some systems. However, this aspect is orthogonal to the focus of this paper, and the techniques developed in the literature \cite{diffmax}, \cite{locbui} to address this problem is complementary to our TCP-aware backpressure framework.}, where the routing and intra-node scheduling algorithm operates as seen in Fig.~\ref{fig:protocol_stack}. The algorithm requires each node to know the queue size of their neighbors. To achieve this, each node $i$ transmits a message containing the size of its per-flow queue sizes; $U_{i}^{s}$ at time $t$. These messages are piggy-backed to data packets. If there is no data transmission for some time duration, our algorithm uses independent control packets to exchange the queue size information. The transmitted message is overheard by all nodes in the neighborhood. 
The queue size information is extracted from the overheard messages and recorded for future decisions.

At node $i$ at time $t$, the queue backlog difference is calculated according to Eq.~(\ref{eq:per_flow_difference}). Note that, although the algorithm exactly knows $U_{i}^{s}(t)$ at time $t$, it is difficult to exactly know $U_{j}^{s}(t)$ at time $t$. Therefore, the most recent report (until time $t$) of the size of $\Uset_{j}^{s}$ is used instead of $U_{j}^{s}(t)$.
When a transmission opportunity for link $(i,j)$ arises using inter-node scheduling algorithm, a packet from flow $s_{i,j}^{*}(t)$ is selected and passed to the MAC layer for transmission.

\subsubsection{Network Coding}
Out of order delivery, high jitter in RTTs, and packet losses over wireless links are among the challenges when backpressure and TCP are combined. We address these challenges by employing network coding \cite{NC_meets_TCP}, \cite{multipath_tcp_toledo}, \cite{i2nc}. This is an effective solution thanks to the properties of network coding such as masking wireless losses and making packets sequence agnostic in terms of packet IDs. We summarize our implementation in the following.

We implement the generation based network coding \cite{practical_NC} at the edge points of the wireless network (\eg access point, base station, proxy, or TCP source itself) as a slim network coding layer (NC layer) above the network layer as shown in Fig.~\ref{fig:protocol_stack}. Note that we do not make any updates to TCP, which makes our approach amenable to practical deployment.

The NC layer at the edge point receives and packetizes the data stream into packets  $\eta_1^{s}, \eta_2^{s}, ...$ of flow $s \in \Sset$. The stream of packets are divided into blocks of size $H_s$, which is set to TCP congestion window size (or its average). The packets within the same block are linearly combined (assuming large enough field size) to generate $H_s$ network coded packets; $a_1^{s} = \alpha_{1,1} \eta_1^{s}$, $a_2^{s} = \alpha_{2,1}\eta_1^{s} + \alpha_{2,2}\eta_2^{s}$, $...$, $a_{H_{s}}^{s} = \alpha_{H_{s},1}\eta_1^{s} + ... + \alpha_{H_{s},H_{s}}\eta_{H_{s}}^{s}$, where $\alpha_{i,j}$, $\forall i,j$ are network coding coefficients from a finite field. Note that network coded packets are generated incrementally to avoid coding delay \cite{practical_NC}, \cite{i2nc}. The NC layer adds network coding header including block ID, packet ID, block size, and coding coefficients. The network coded packets are routed and scheduled by TCP-aware backpressure.

At the receiver node, when the NC layer receives a packet from a new block, it considers the received packet as the first packet in the block. It generates an ACK, sends the ACK back to the NC layer at the edge point, which matches this ACK to packet $\eta_1$, converts this ACK to $\eta_1$'s ACK, and transmits the ACK information to the TCP source. Similarly, ACKs are generated at the receiver side for the second, third, etc. received packets. As long as the NC layer at the receiver transmits ACKs, the TCP clock moves, and the window continues to advance.

The NC layer stores the received network coded packets in a buffer. When the last packet from a block is received, packets are decoded and passed to the application layer. If some packets are lost in the wireless network, the receiver side NC layer makes a request with the block ID and the number of missing packets, and the edge point side NC layer generates additional network coded packets from the requested block, and sends to the receiver. Note that the missing packet IDs are not mentioned in the request, since the network coding makes the packets sequence agnostic in terms of packet IDs.

Network coding makes packets sequence agnostic, which solves out of order delivery problem and eliminates jitter. Network coding also corrects packet losses in the wireless network as explained above. We explain how our system and NC layer reacts to congestion-based losses later in this section.




\subsection{Interaction with TCP}

\subsubsection{Congestion Control}
Now, let us consider the interaction of TCP congestion control and TCP-aware backpressure using well-known classical TCP analysis \cite{twsly_tcp}, \cite{low_tcp}. Using the similar approach as in \cite{twsly_tcp}, \cite{low_tcp}, and as detailed in \cite{tcp_aware_bp_tech_rep}, we find the steady state TCP throughput for flow $s$ as; $x_{s}^{2} = \frac{(1-q_{o(s)}^{s})}{T_{s}^{3}q_{o(s)}^{s}}$, where $q_{o(s)}^{s}$ is the buffer overflow probability at the TCP source/edge node $o(s)$, and $T_s$ is constant RTT.\footnote{\scriptsize The constant RTT is a common assumption in classical TCP analysis \cite{twsly_tcp}, \cite{low_tcp}, and also valid in our setup thanks to employing network coding, which reduces jitter in RTT and makes constant RTT assumption valid.}

Note that the steady state TCP throughput depends on the buffer overflow probability only at the source/edge node different from \cite{twsly_tcp}, \cite{low_tcp}, where TCP throughput depends on the buffer overflow probability over all nodes over the path of TCP flow.\footnote{\scriptsize Note that steady state TCP throughput does not depend on packet trapping events thanks to employing Eq.~(\ref{eq:per_flow_difference}). This does not hold for classical backpressure, because some packets may be trapped in buffers, which reduces TCP throughput, and should be taken into account in the steady state TCP throughput analysis.} The reason is that congestion in the wireless network is controlled by TCP-aware backpressure, and we do not expect losses due to congestion (buffer overflow) at the intermediate nodes. In particular, as TCP-aware backpressure makes transmission decisions based on queue backlog differences according to Eq.~(\ref{eq:per_flow_difference}), it would not transmit packets if the next hop queue is congested. Therefore, congestion-based losses only occur at the source/edge node. In our implementation, if the buffer at the source/edge node is congested, than a packet from the flow which has the largest queue size is dropped. This congestion-based loss information is passed to the NC layer. The NC layer creates a loss event by not masking the dropped packet so that TCP can detect the congestion-based loss event and back-off. 



\subsubsection{Selection of $K$}
TCP-aware backpressure uses queue backlog difference in Eq.~(\ref{eq:per_flow_difference}), which depends on $K$, to make routing and scheduling decisions. As noted in Section~\ref{sec:opt}, the selection of $K$ is crucial in practice to make TCP and backpressure fully comply.

In particular, if $K$ is selected too small, the number of packets that are trapped in the buffers, \ie the number of packets that do not get transmission opportunity, increases. This reduces TCP throughput. On the other hand, if $K$ is too large, TCP-aware backpressure may not exploit the throughput improvement benefit of backpressure routing and scheduling as the ability of identifying good routing and scheduling policies reduces with large $K$ values.


Our intuition is that flows passing through node $i$, \ie $s \in \Sset_{i}$, should share the available buffer fairly. Assume that $B_i$ is the available buffer size at node $i$. In order to give transmission opportunity to all TCP flows and provide some level of fairness across the competing TCP flows, we set $K = B_{i} / |\Sset_{i}|$ at node $i$. In this setting, if per-flow queue sizes are smaller than $K$, it is highly possible that packets from all TCP flows are transmitted. On the other hand, if some per-flow queue sizes are larger than $K$, packets from the flows with smaller queue sizes may still be trapped in the buffers. However, in this case, since the total buffer occupancy is large, buffer overflow probability at the source/edge node increases. Upon buffer overflow, the TCP flow with larger queue size reduces its rate (since upon congestion a packet from the largest per-flow queue is dropped). This reduces the queue sizes, and packets from all flows could be transmitted again.

{\em Example 2 - continued:} Let us consider again Fig.~\ref{fig:intro_example}(b). If the queue occupancies are $U_{I}^{1}(t) = 7$, $U_{I}^{2}(t) = 12$, and $K=10$, packets only from the second flow are transmitted. Since $K=10$ and we set $K = B_{I} / |\Sset_{I}|$, and $|\Sset_{I}|=2$, the buffer size is $B_{I}=20$. The total queue occupancy is  $U_{I}^{1}(t)+U_{I}^{2}(t)=19$. This means that the buffer at node $I$ is about to overflow, which will lead to back-off for the second flow (since a packet from the largest queue will be dropped). Thus, the TCP rate and queue size of the second flow will reduce, and the first flow will get transmission opportunity. 
\hfill $\blacksquare$

We have observed through simulations that TCP-aware backpressure, when $K$ is set to $B_{i} / |\Sset_{i}|$, significantly reduces the number of the trapped packets in the buffers. Yet, very few packets may still be trapped. Such packets are easily masked thanks to error correction capabilities of network coding. Note that network coding does not help if large number of packets are trapped in the buffers (\eg when $K$ is selected too small), as large number of trapped packets increases end-to-end delay too much, which leads to multiple timeouts and reduces TCP throughput.

\section{Performance Evaluation}\label{sec:performance}
We simulate our scheme, TCP-aware backpressure (TCP-aware BP) as well as classical backpressure (classical BP), in ns-2 \cite{ns2}. The simulation results; (i) confirm the mismatch of TCP and classical BP, (ii) show that TCP-aware BP is compatible with TCP, and significantly improves throughput as compared to existing routing schemes such as Ad-hoc On-Demand Distance Vector (AODV) \cite{aodv}, (iii) demonstrate that TCP-aware BP provides fairness across competing TCP flows. Next, we present the simulator setup and results in detail.

\begin{figure*}[t!]
\vspace{-5pt}
\centering
\subfigure[\scriptsize  Tree topology]{ \label{fig:intro_example_a} \scalebox{.40}{\includegraphics[bb=0 0 246 164]{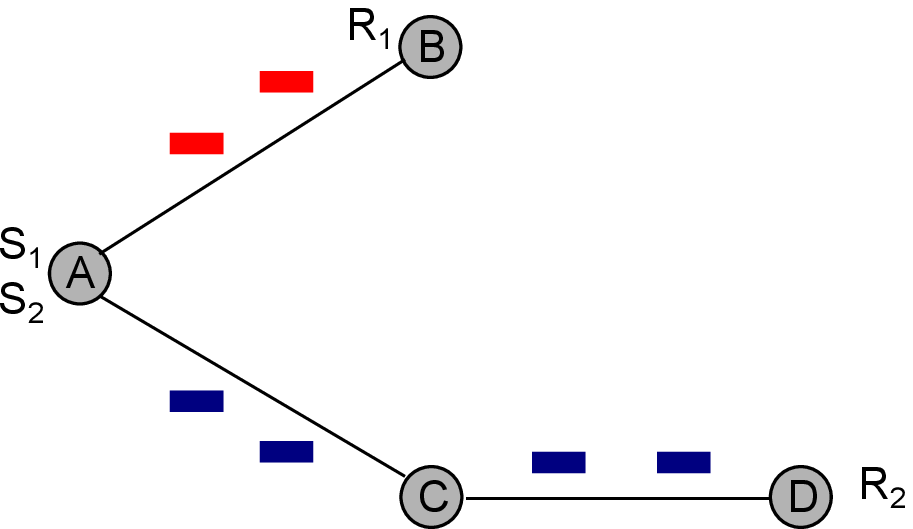}} }
\subfigure[\scriptsize Diamond topology]{ \label{fig:intro_example_a} \scalebox{.45}{\includegraphics[bb=0 0 246 164]{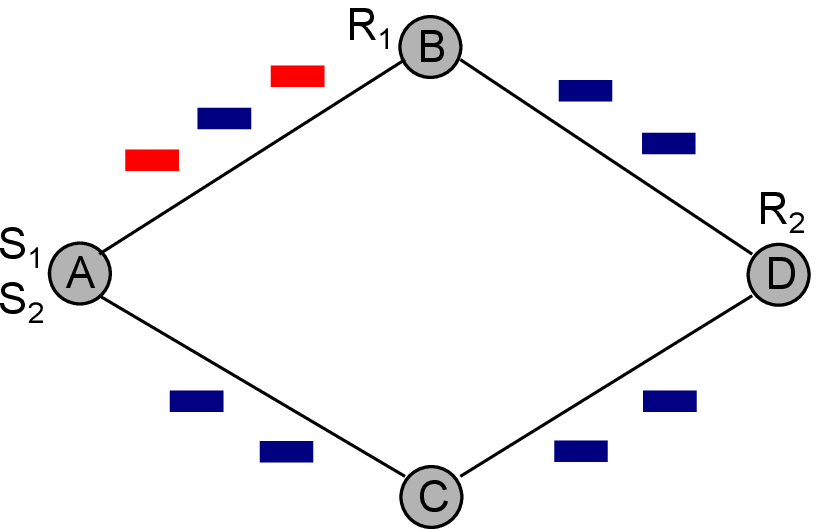}} }
\subfigure[\scriptsize Grid topology]{ \label{fig:intro_example_b} \scalebox{.35}{\includegraphics[bb=0 0 299 179]{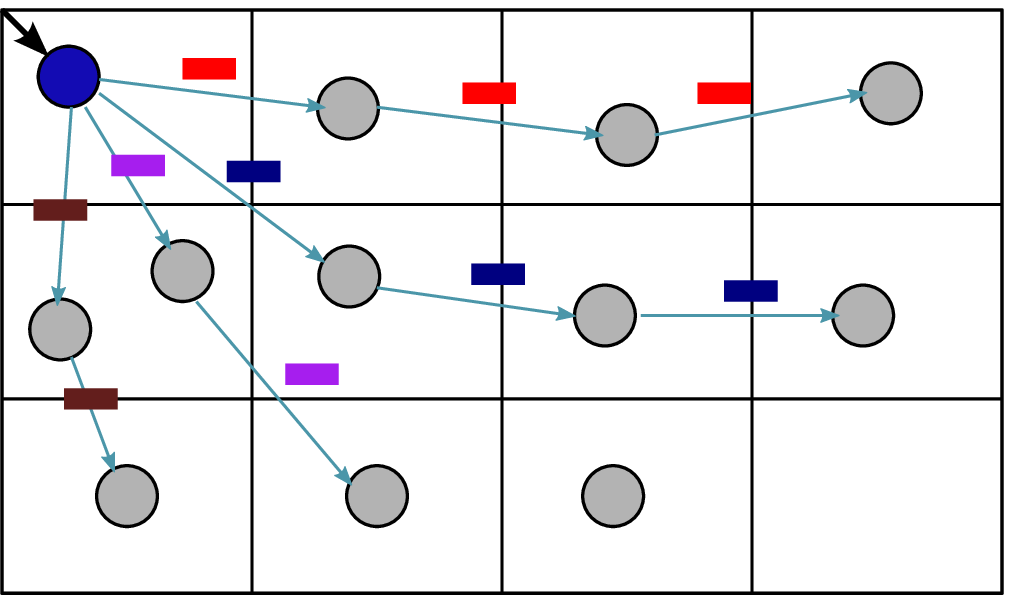}} }
\vspace{-5pt}
\caption{\scriptsize Topologies used in simulations; (a) tree topology, (b) diamond topology, (c) grid topology.}
\vspace{-15pt}
\label{fig:topologies}
\end{figure*}


\subsection{Simulation Setup}
We consider three topologies: a tree topology, a diamond topology, and a grid topology shown in Fig.~\ref{fig:topologies}. The nodes are placed over $500m \times 500m$ terrain, and $S_1$, $S_2$ and $R_1$, $R_2$ are possible source-receiver pairs in the tree and diamond topologies. In the grid topology, $4 \times 3$ cells are placed over a $800m \times 600m$ terrain. A gateway, which is connected to the Internet, passes flows to nodes. Each node communicates with other nodes in its cell or neighboring cells, and there are $12$ nodes randomly placed to the cells.

We consider FTP/TCP traffic, and employ TCP-SACK and TCP-Vegas in our simulations. TCP flows start at random times within the first $5sec$ of the simulation and are on until the end of the simulation which is $200sec$. IEEE 802.11b is used in the MAC layer. In terms of wireless channel, we simulated the two-ray path loss model and a Rayleigh fading channel with average loss rates $0, 20, 30, 40, 50 \%$.
Channel capacity is $2Mbps$, the buffer size at each node is set to $100$ packets, packet sizes are set to $1000B$. We have repeated each $200sec$ simulation for 10 seeds.

We compare our scheme, TCP-aware BP, to the classical BP and AODV. For fair comparison, we employ the network coding mechanism explained in Section~\ref{sec:algs} in the classical BP as well as in AODV. The comparisons are in terms of per-flow and total transport level throughput (added over all flows) as well as fairness. For the fairness calculation, we use Jain's fairness index \cite{fairness_index}: $F = \frac{(\sum_{s \in \Sset} \bar{x}_s)^2}{|\Sset|(\sum_{s \in \Sset} (\bar{x}_s)^2)}$, where $\Sset$ is the set of flows and $\bar{x}_s$ is the average throughput of flow $s$.


\subsection{Simulation Results}

\begin{figure}[t!]
\vspace{-0pt}
\begin{center}
\subfigure[\scriptsize{TCP-Aware BP with TCP-SACK}]{{\includegraphics[width=4cm]{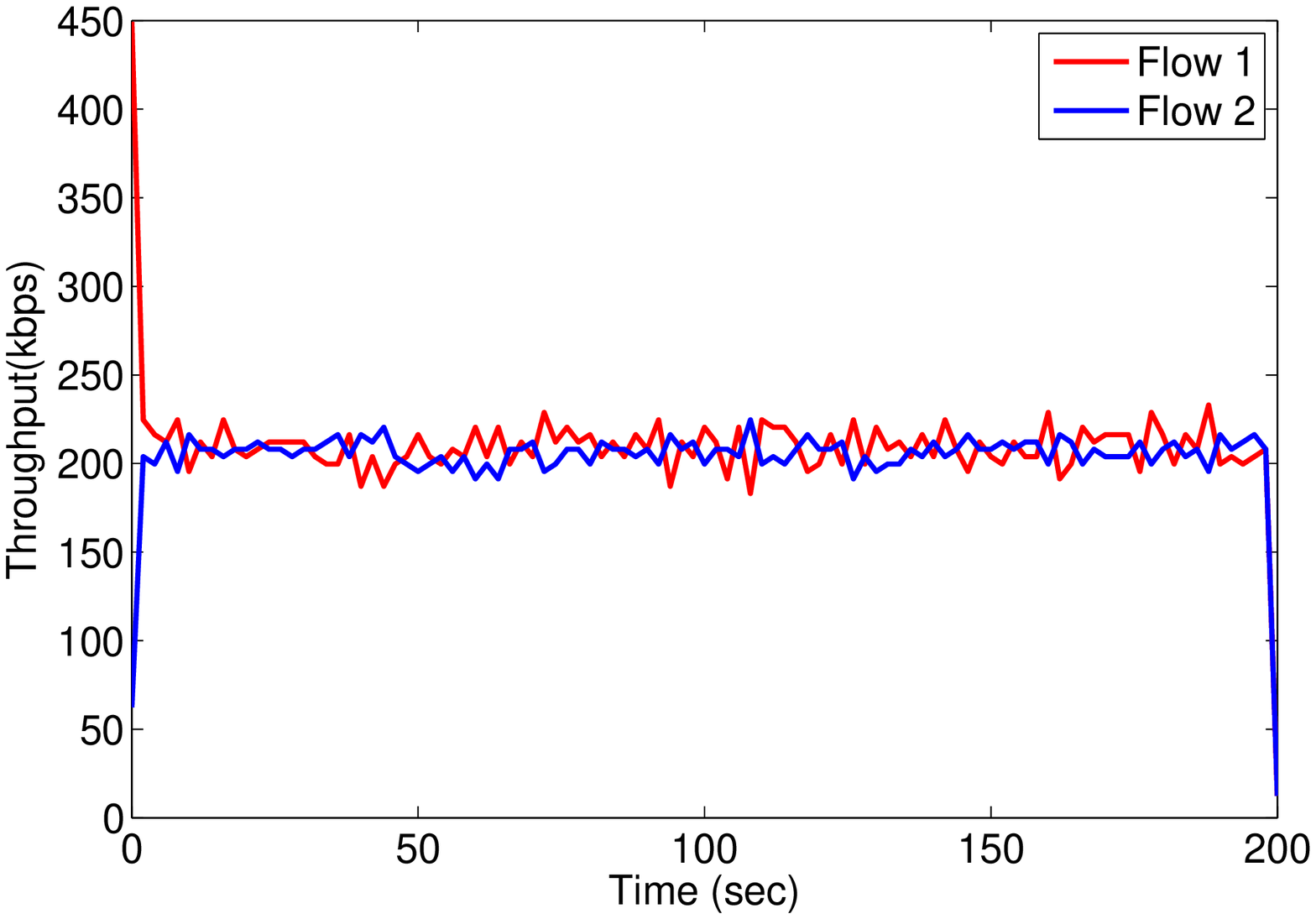}}}
\subfigure[\scriptsize{BP with TCP-SACK}]{{\includegraphics[width=4cm]{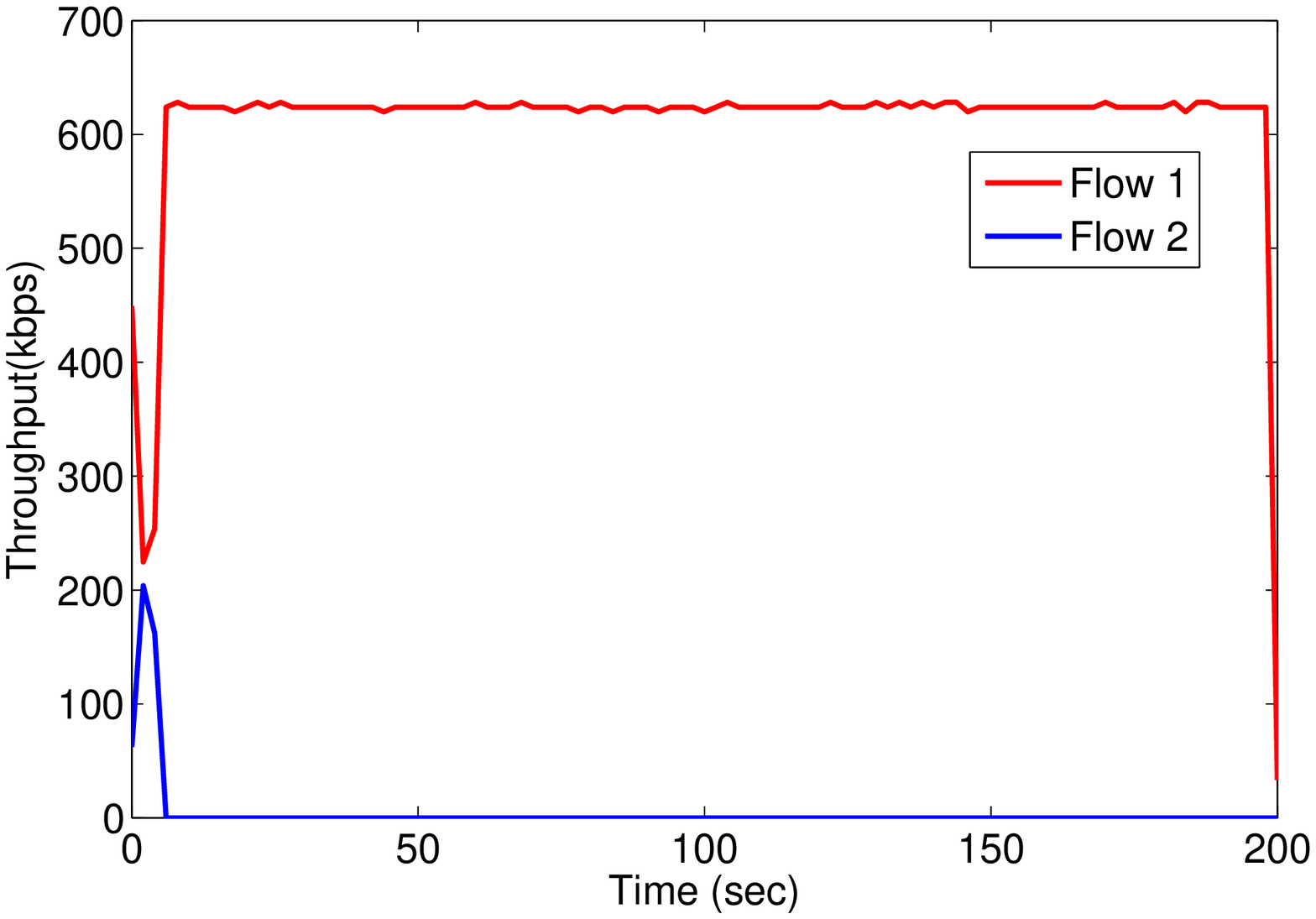}}} \hspace{-0pt} \\
\subfigure[\scriptsize{TCP-Aware BP with TCP-Vegas}]{{\includegraphics[width=4cm]{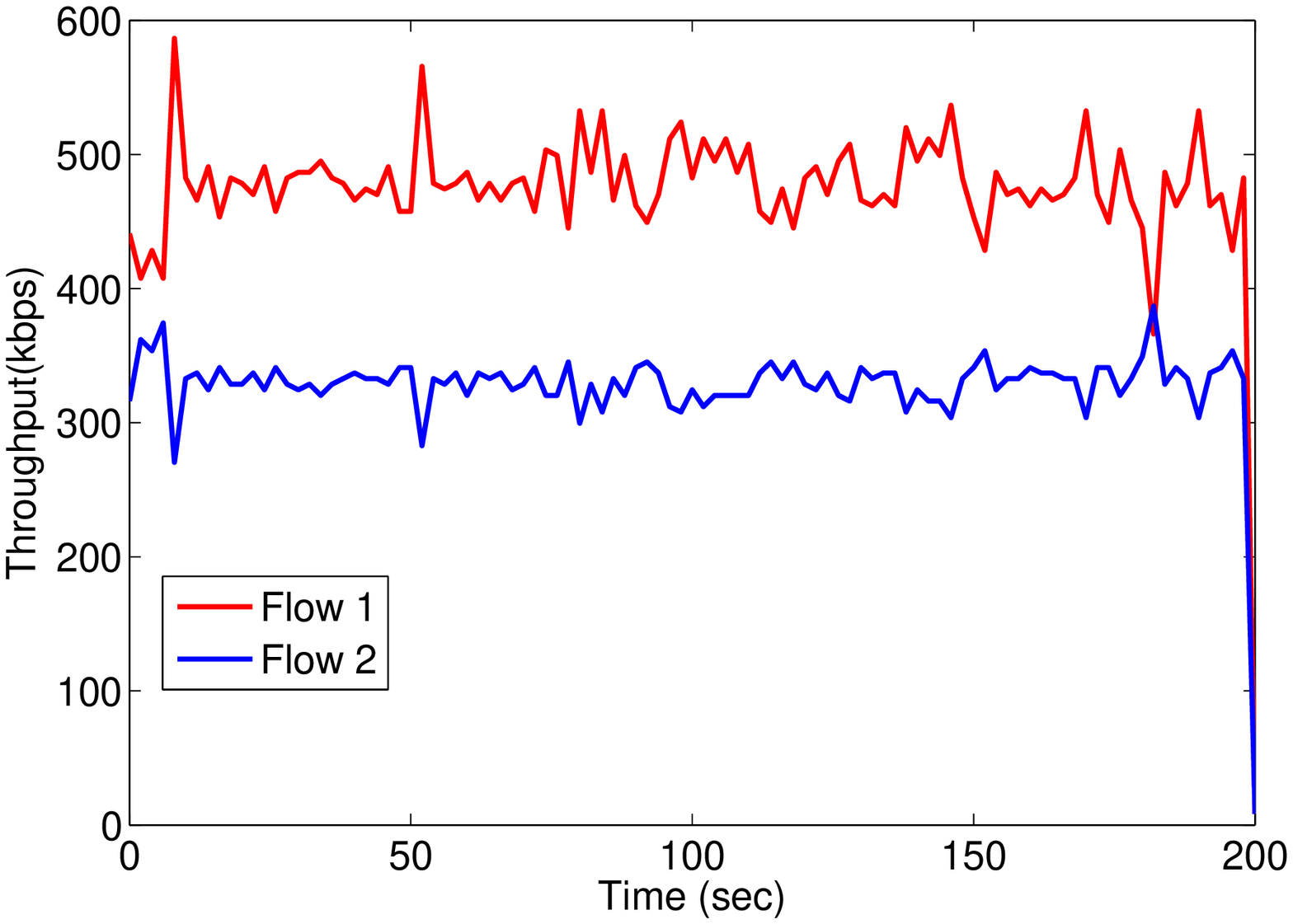}}}
\subfigure[\scriptsize{BP with TCP-Vegas}]{{\includegraphics[width=4cm]{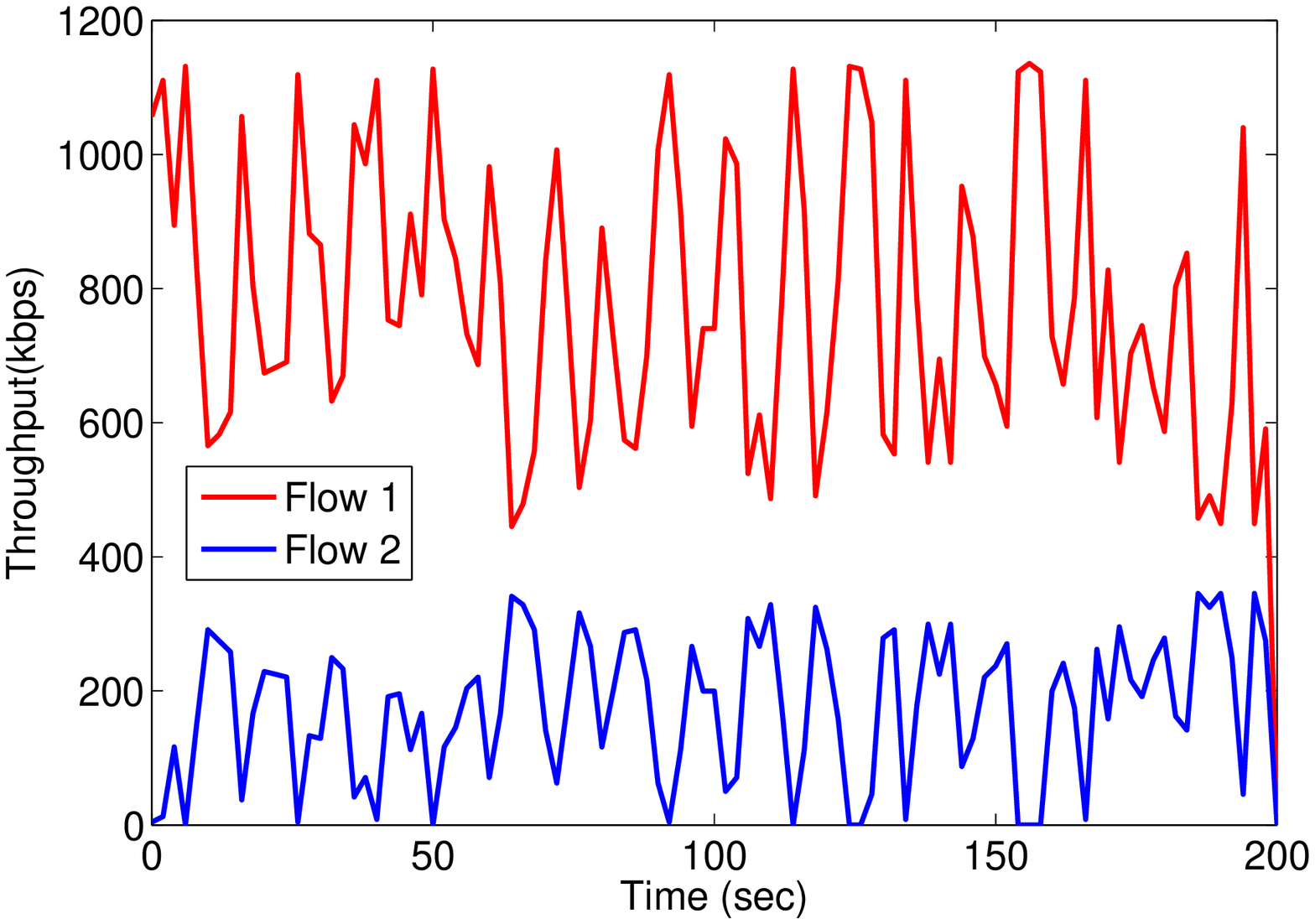}}} \hspace{-0pt}
\end{center}
\begin{center}
\vspace{-5pt}
\caption{\label{fig:tree_thrpt_time_results} \scriptsize  Throughput vs. time in the tree topology for TCP-SACK and TCP-Vegas. There are two flows; Flow 1 is transmitted from node $A$ to node $B$, and Flow 2 is transmitted from node $A$ to node $D$. The links are not lossy.
}
\vspace{-15pt}
\end{center}
\vspace{-10pt}
\end{figure}

\begin{figure}[t!]
\vspace{-0pt}
\begin{center}
\subfigure[\scriptsize TCP-Aware BP with TCP-SACK]{{\includegraphics[width=4cm]{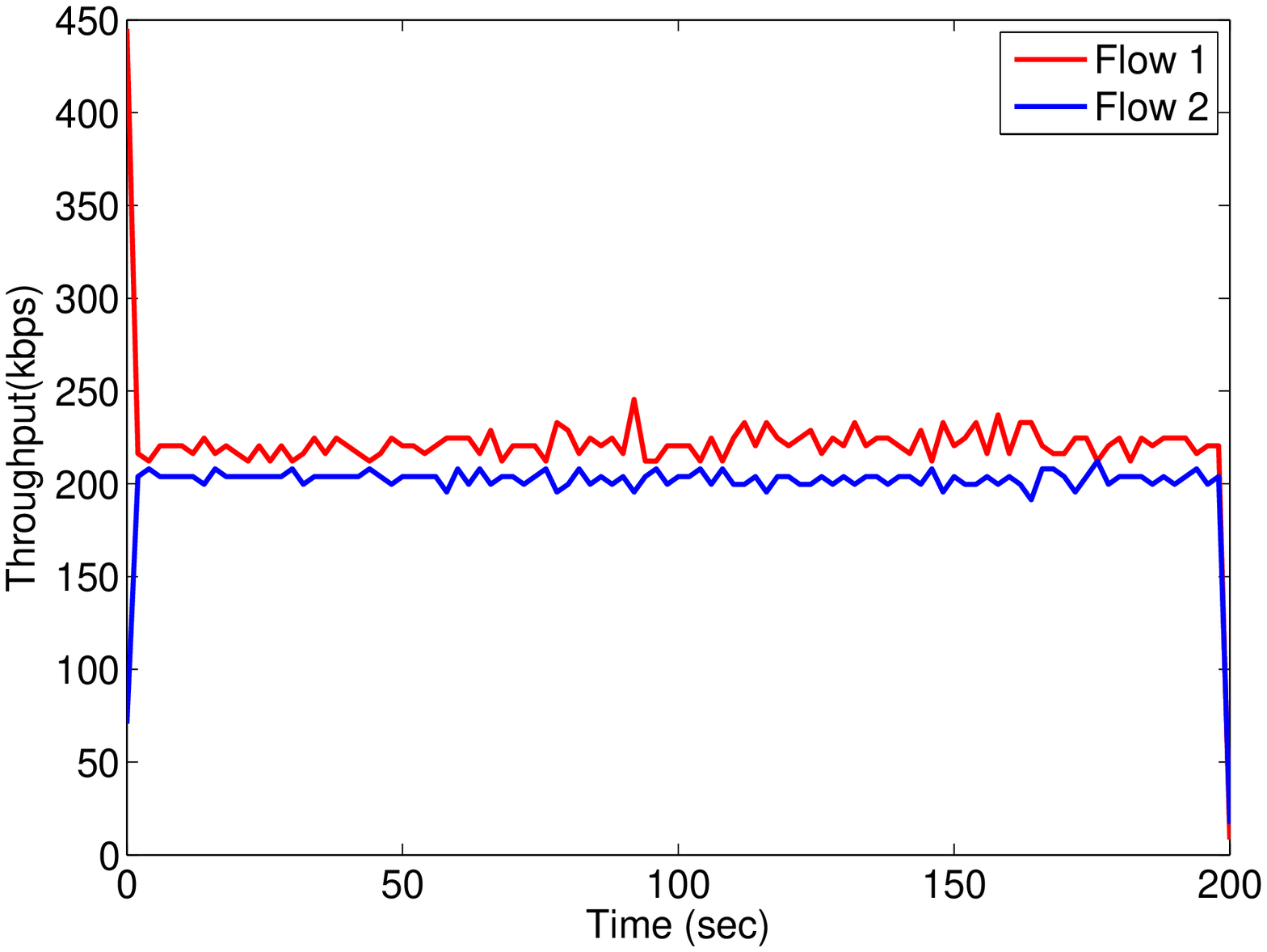}}}
\subfigure[\scriptsize BP with TCP-SACK]{{\includegraphics[width=4cm]{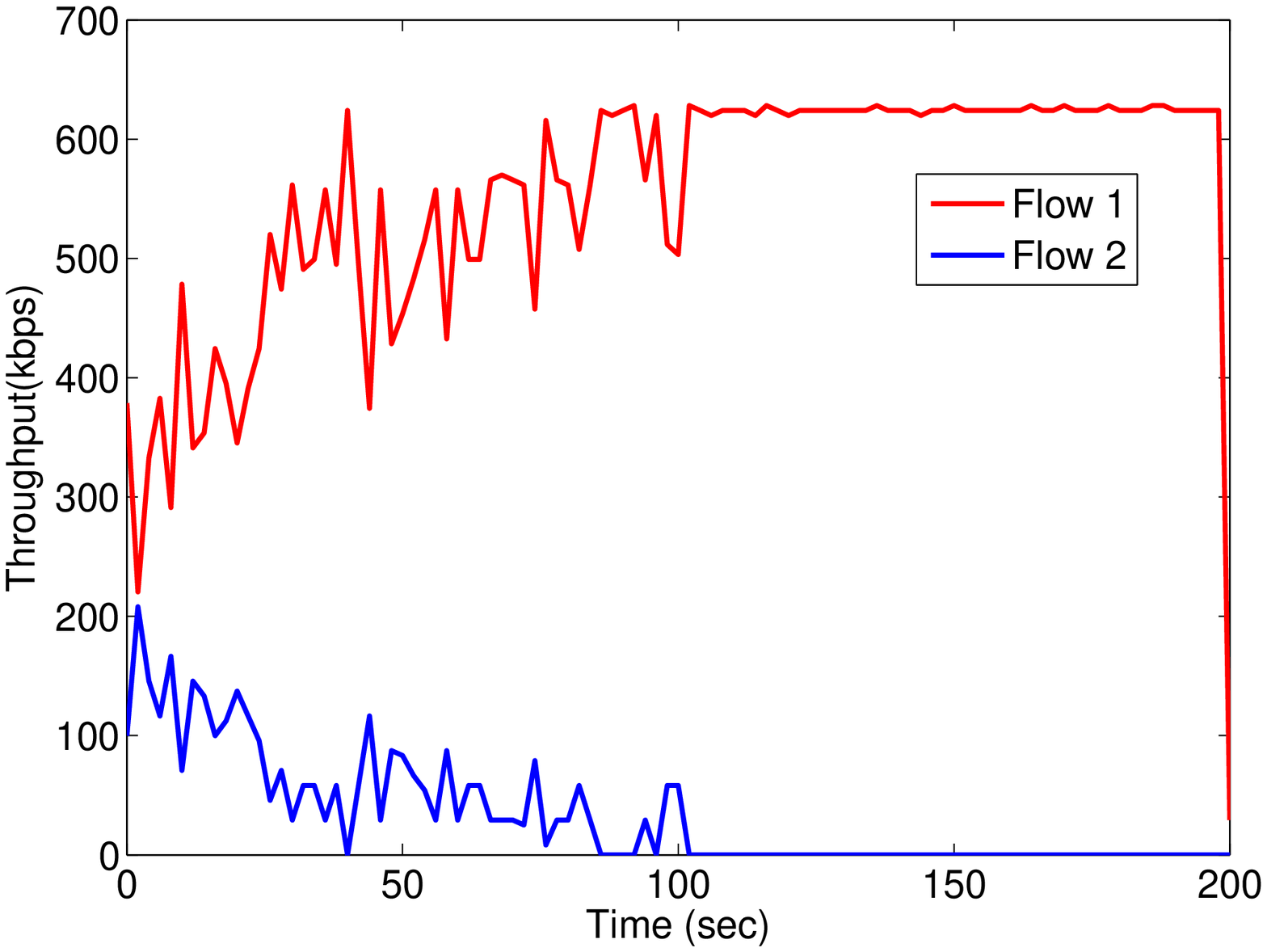}}} \hspace{-0pt} \\
\subfigure[\scriptsize TCP-Aware BP with TCP-Vegas]{{\includegraphics[width=4cm]{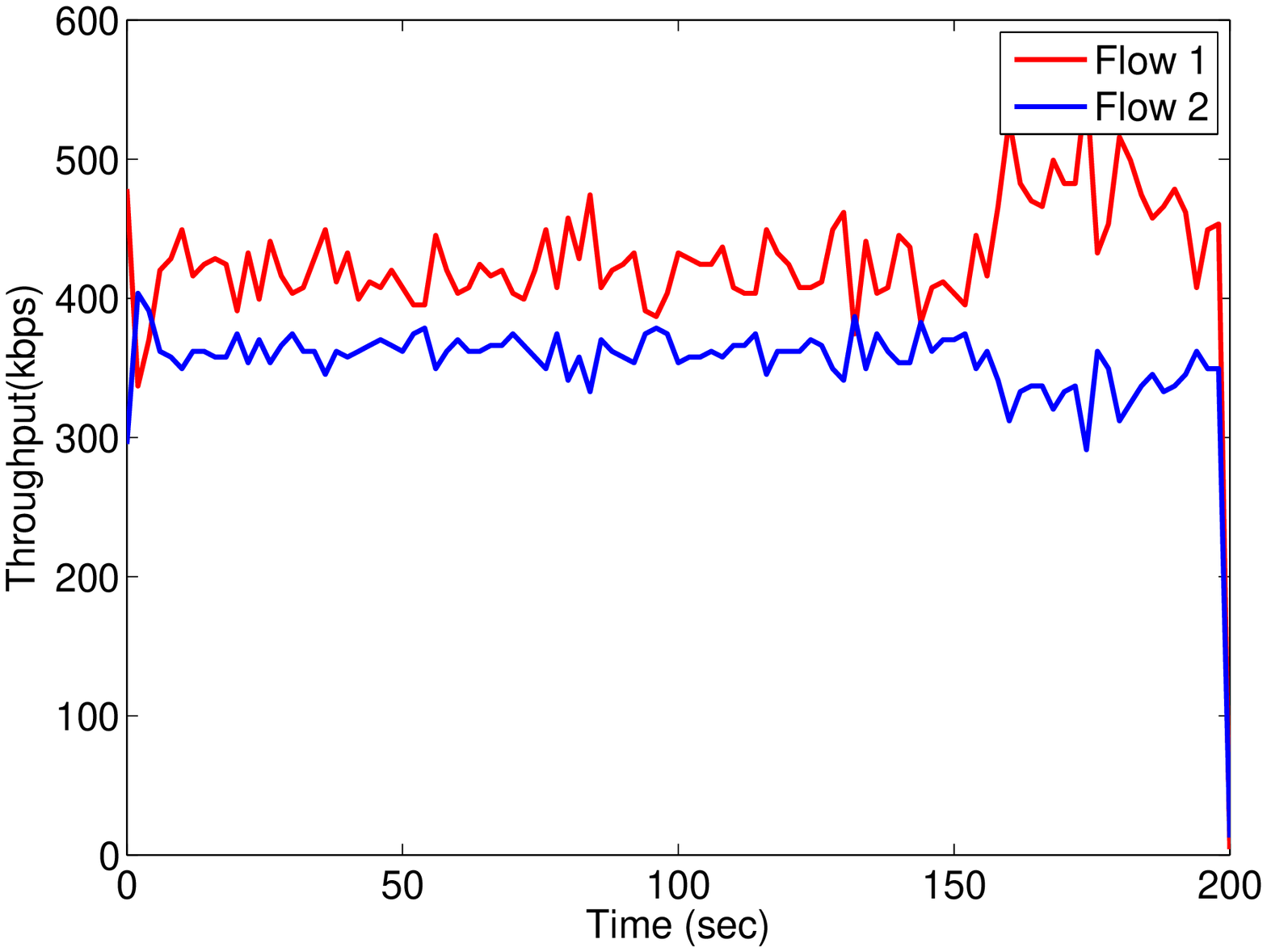}}}
\subfigure[\scriptsize BP with TCP-Vegas]{{\includegraphics[width=4cm]{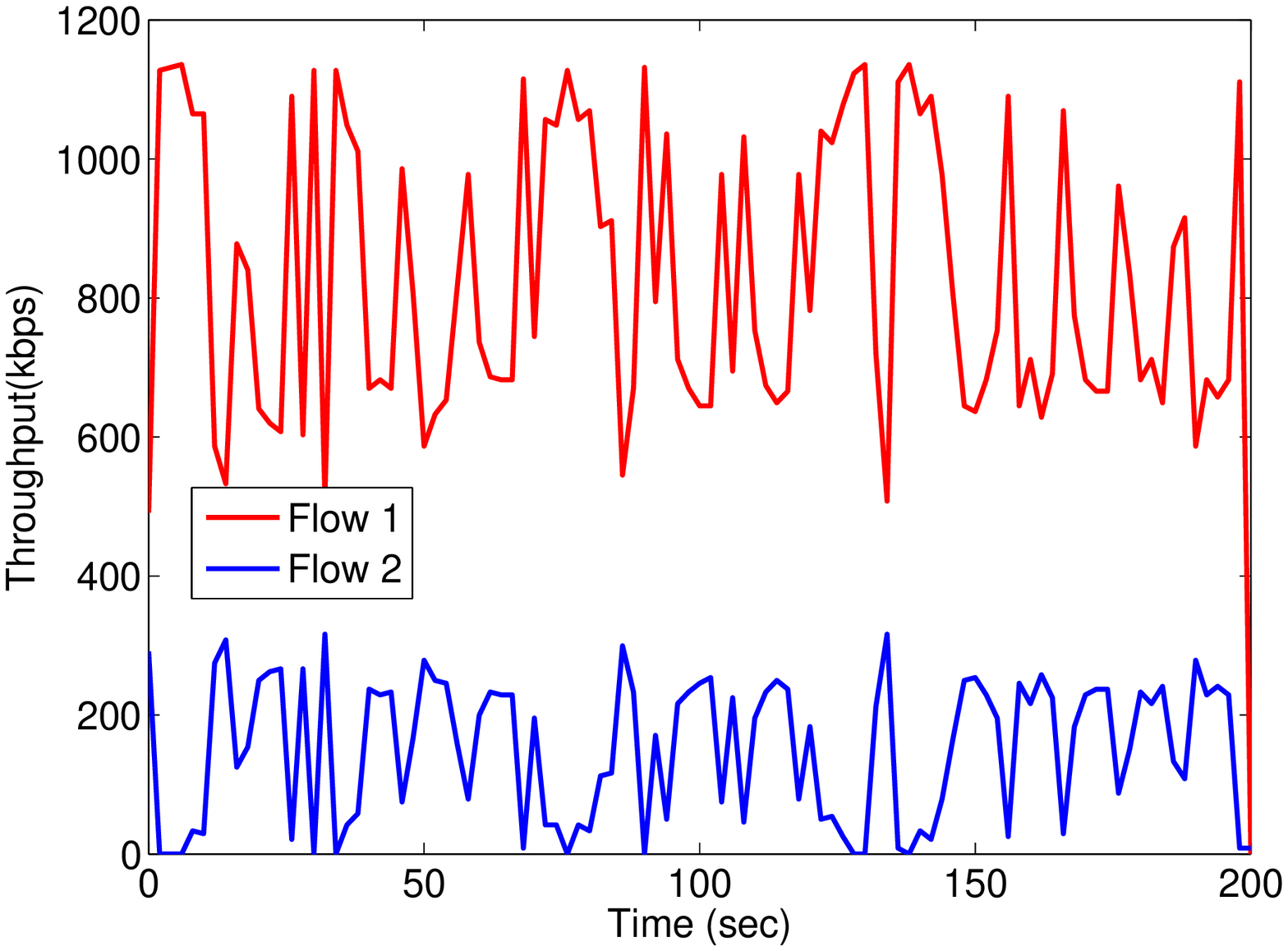}}} \hspace{-0pt}
\end{center}
\begin{center}
\vspace{-5pt}
\caption{\label{fig:diamond_thrpt_time_results} \scriptsize Throughput vs. time in the diamond topology for TCP-SACK and TCP-Vegas. There are two flows; Flow 1 is transmitted from node $A$ to node $B$, and Flow 2 is transmitted from node $A$ to node $D$. The links are not lossy.
}
\vspace{-15pt}
\end{center}
\end{figure}

Fig.~\ref{fig:tree_thrpt_time_results} shows throughput vs. time graphs for TCP-aware BP and classical BP. 
There are two flows; Flow 1 is transmitted from node $A$ to node $B$, and Flow 2 is transmitted from node $A$ to node $D$. The links are not lossy. Fig.~\ref{fig:tree_thrpt_time_results}(a) and (b) are the results for TCP-SACK, while Fig.~\ref{fig:tree_thrpt_time_results}(c) and (d) are for TCP-Vegas. Fig.~\ref{fig:tree_thrpt_time_results}(b) shows that while Flow 1 is able to transmit, Flow 2 does not get any chance for transmission in classical BP due to the mismatch between congestion window size update mechanism of TCP and queue size-based routing and scheduling of backpressure. On the other hand, in TCP-aware BP, both flows get chance for transmission. In particular, Flow 1 and Flow 2 achieves average throughput of $205.76 kbps$ and $203.36 kbps$, respectively. Fig.~\ref{fig:tree_thrpt_time_results}(c) and (d) show throughput vs. time graphs of TCP-aware BP and classical BP for TCP-Vegas. Although classical BP performs better in TCP-Vegas than in TCP-SACK due to the delay based mechanism of TCP-Vegas, its performance is still quite poor as the throughput of Flow 2 frequently goes to 0 as seen in Fig.~\ref{fig:tree_thrpt_time_results}(d). On the other hand, TCP-aware BP improves throughput of both flows as seen in Fig.~\ref{fig:tree_thrpt_time_results}(c), where Flow 1 and Flow 2 achieve $469.36 kbps$ and $324.64 kbps$, respectively. The similar results are presented in Fig.~\ref{fig:diamond_thrpt_time_results} for the diamond topology.

\begin{figure}[t!]
\vspace{-0pt}
\begin{center}
\subfigure[\scriptsize Throughput]{{\includegraphics[width=4cm]{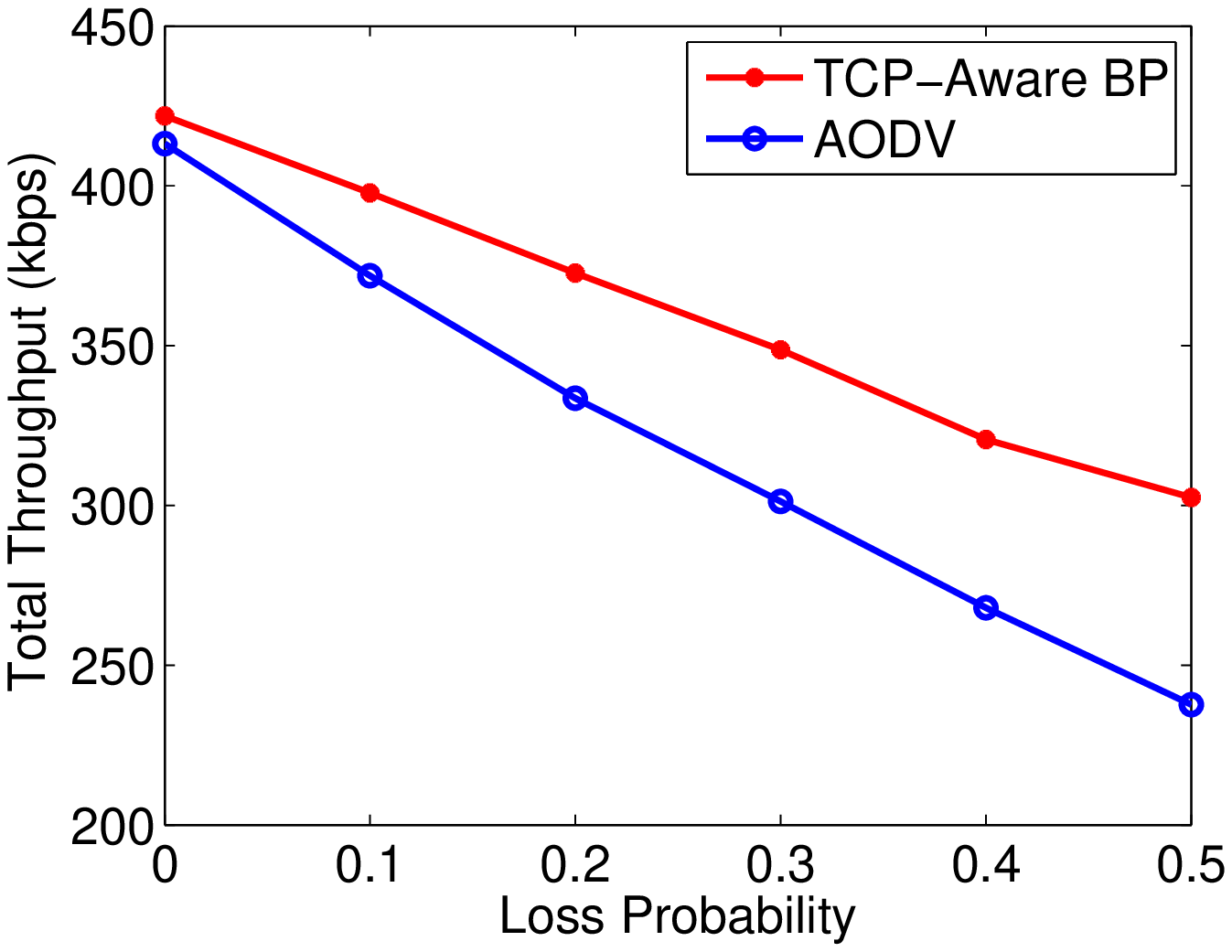}}}
\subfigure[\scriptsize Fairness]{{\includegraphics[width=4cm]{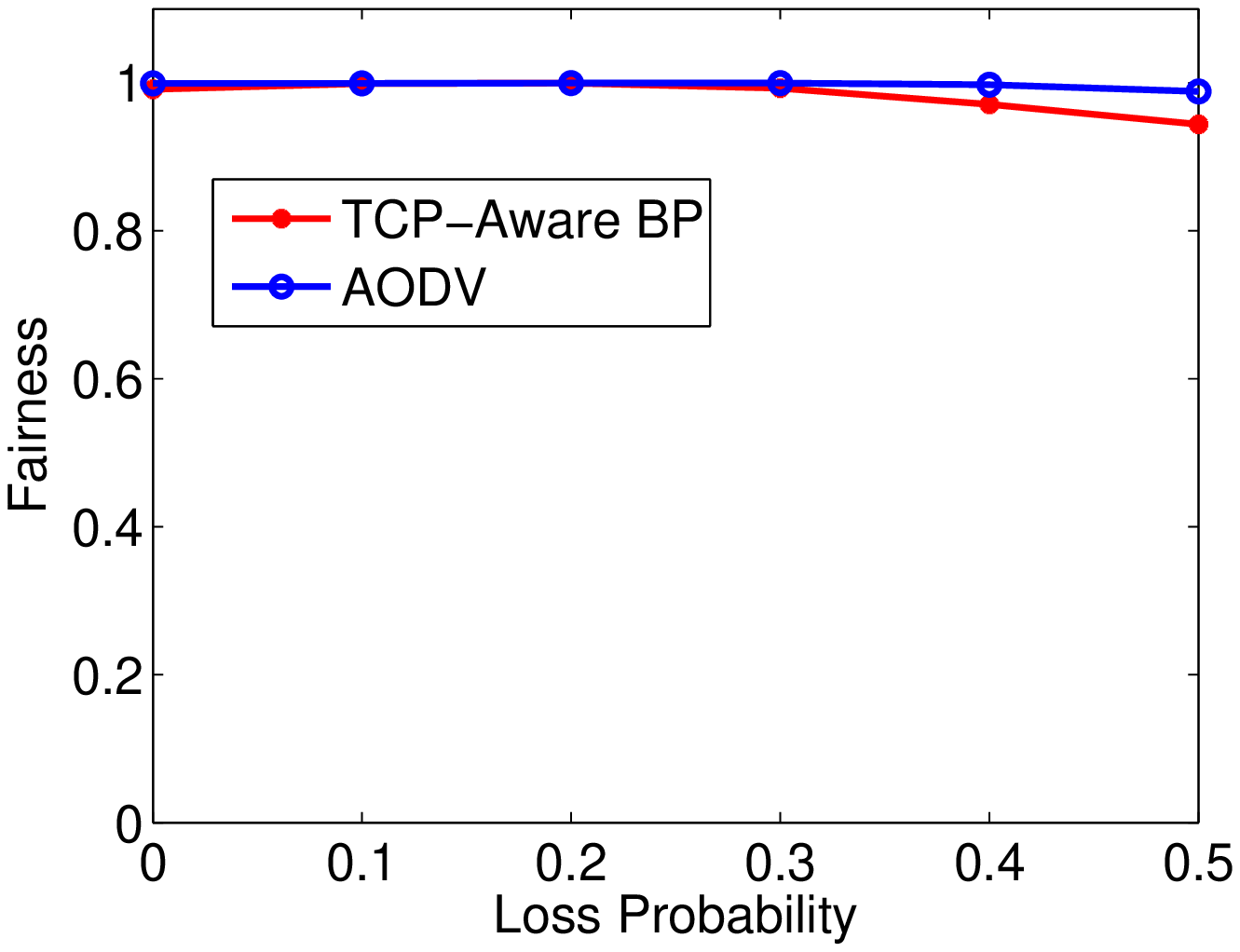}}}
\end{center}
\begin{center}
\vspace{-5pt}
\caption{\label{fig:diamond_thrpt_vs_loss_sack} \scriptsize Throughput and fairness vs. average packet loss rate for TCP-aware BP and AODV in the diamond topology. There are two TCP flows transmitted from node $A$ to $B$ (Flow 1) and $A$ to $D$ (Flow 2). The link $A-B$ is a lossy link. The version of TCP is TCP-SACK.}
\end{center}
\vspace{-15pt}
\end{figure}

Fig.~\ref{fig:diamond_thrpt_vs_loss_sack} demonstrates throughput and fairness vs. average loss rate results of TCP-aware BP and AODV in the diamond topology. There are two flows transmitted from node $A$ to $B$ (Flow 1) and $A$ to $D$ (Flow 2). The link $A-B$ is a lossy link. The version of TCP is TCP-SACK. 
Fig.~\ref{fig:diamond_thrpt_vs_loss_sack}(a) shows that TCP-aware BP improves throughput significantly as compared to AODV thanks to adaptive routing and scheduling. The throughput improvement of TCP-aware BP as compared to AODV increases as loss probability increases thanks to loss-aware routing and scheduling mechanism of TCP-aware BP. Moreover, Fig.~\ref{fig:diamond_thrpt_vs_loss_sack}(b) shows that the fairness index is close to $F=1$ (note that $F=1$ is the highest possible fairness index) when TCP-aware BP is employed. This means that both TCP flows are able to survive in TCP-aware BP. Note that the fairness index of TCP-aware BP is 0.94, while the fairness index of AODV is 0.98 when the packet loss probability is 0.5. This is due to the fact that TCP-aware BP exploits loss-free links better, and slightly favors the flows transmitted over such links. However, the throughput improvement of both flows as compared to AODV is higher. In particular, TCP-aware BP improves throughput as compared to AODV by \%10 and \%40 for the first and second flows, respectively. These results confirm the compatibility of TCP and TCP-aware BP.

\begin{figure}[t!]
\vspace{-0pt}
\begin{center}
\subfigure[\scriptsize TCP-Aware BP with TCP-SACK]{{\includegraphics[width=4cm]{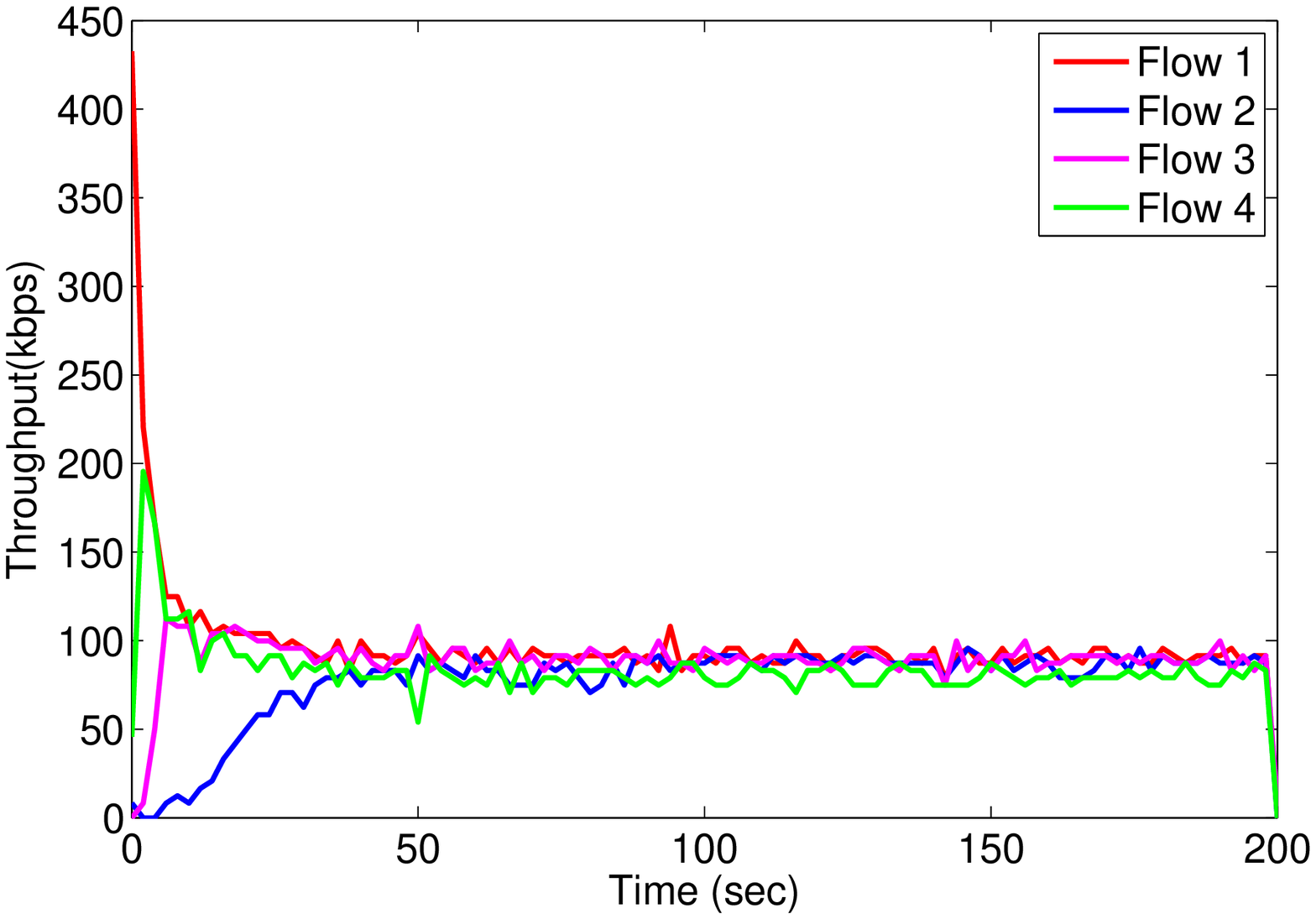}}}
\subfigure[\scriptsize BP with TCP-SACK]{{\includegraphics[width=4cm]{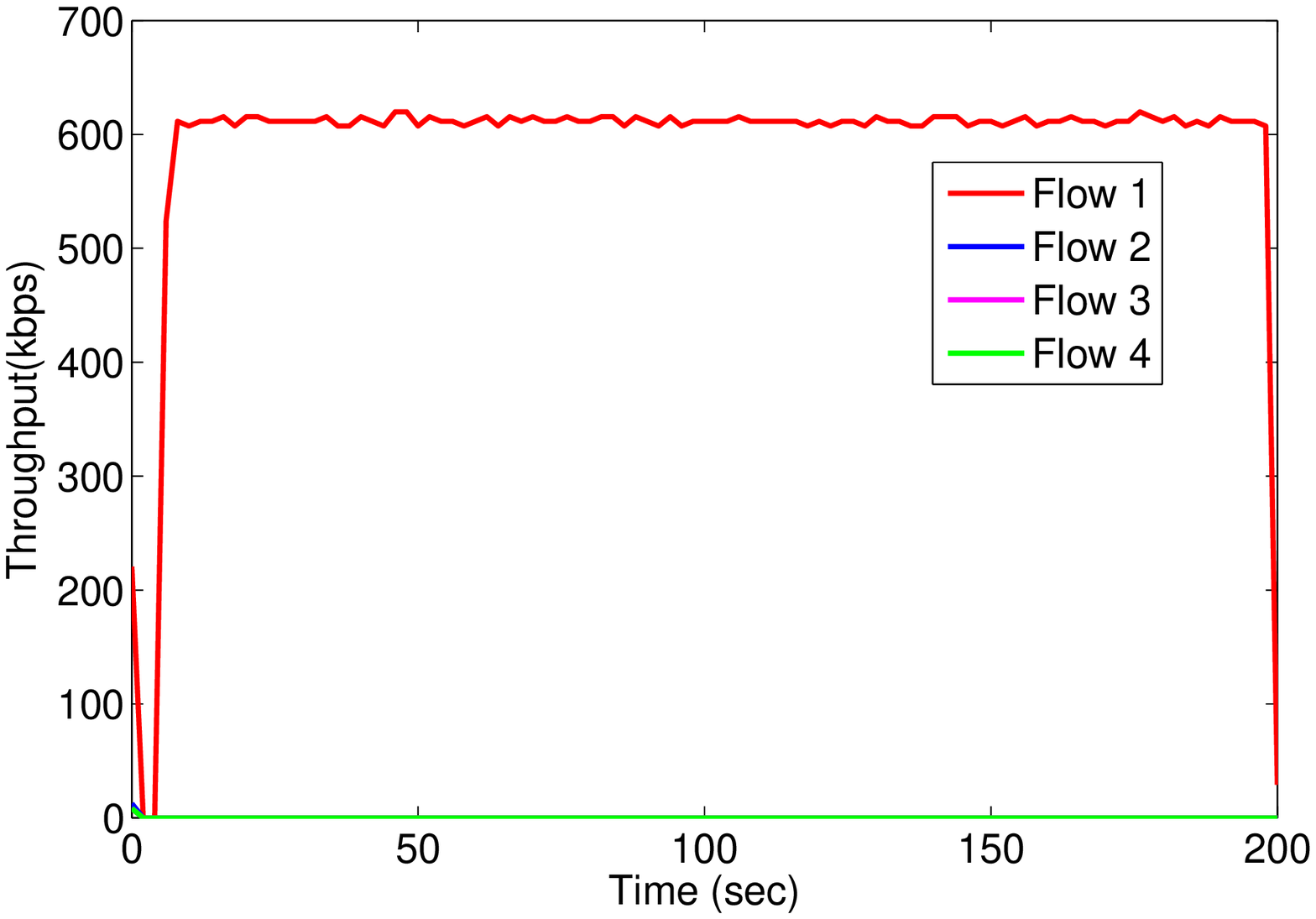}}} \hspace{-0pt} \\
\subfigure[\scriptsize TCP-Aware BP with TCP-Vegas]{{\includegraphics[width=4cm]{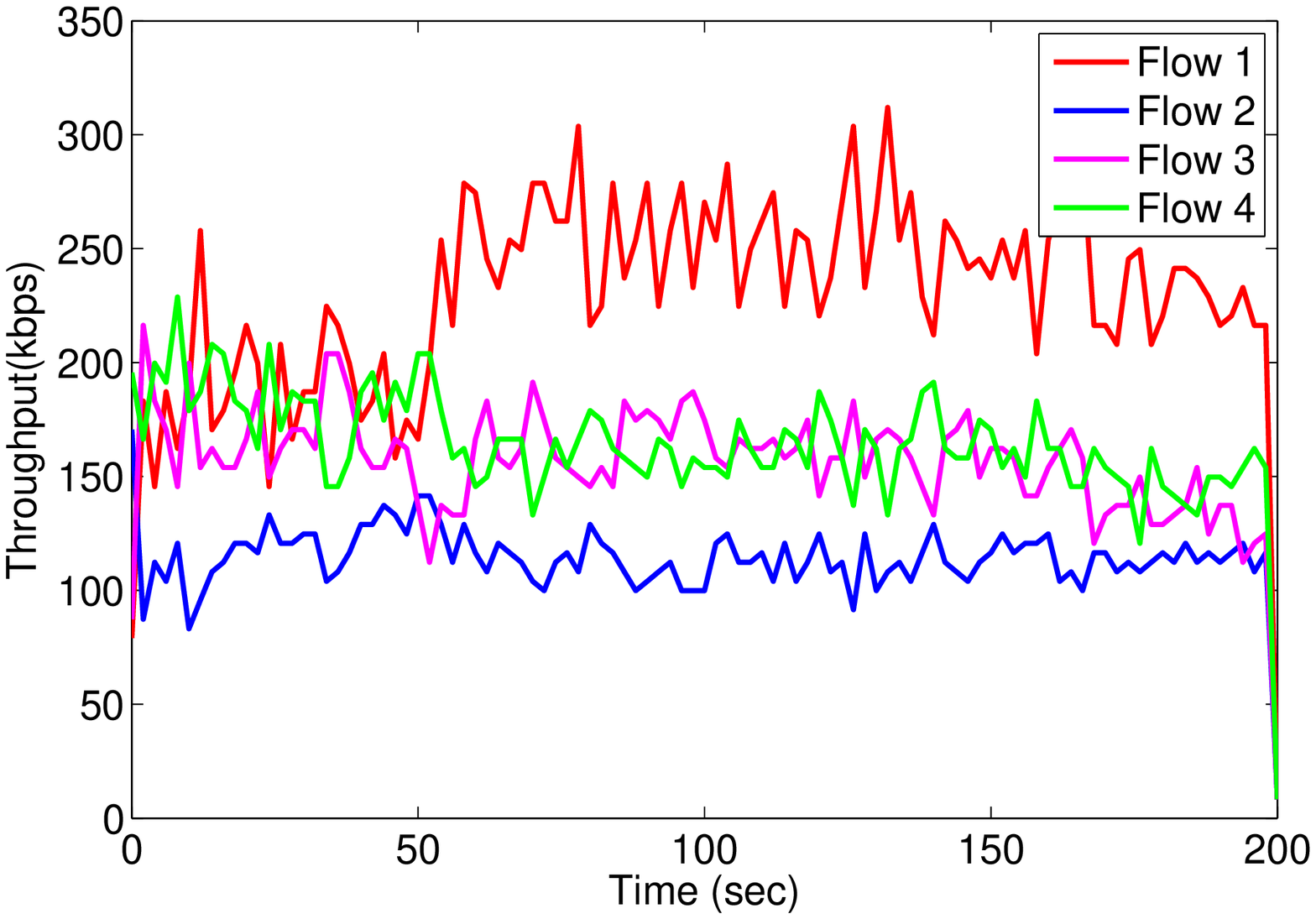}}}
\subfigure[\scriptsize BP with TCP-Vegas]{{\includegraphics[width=4cm]{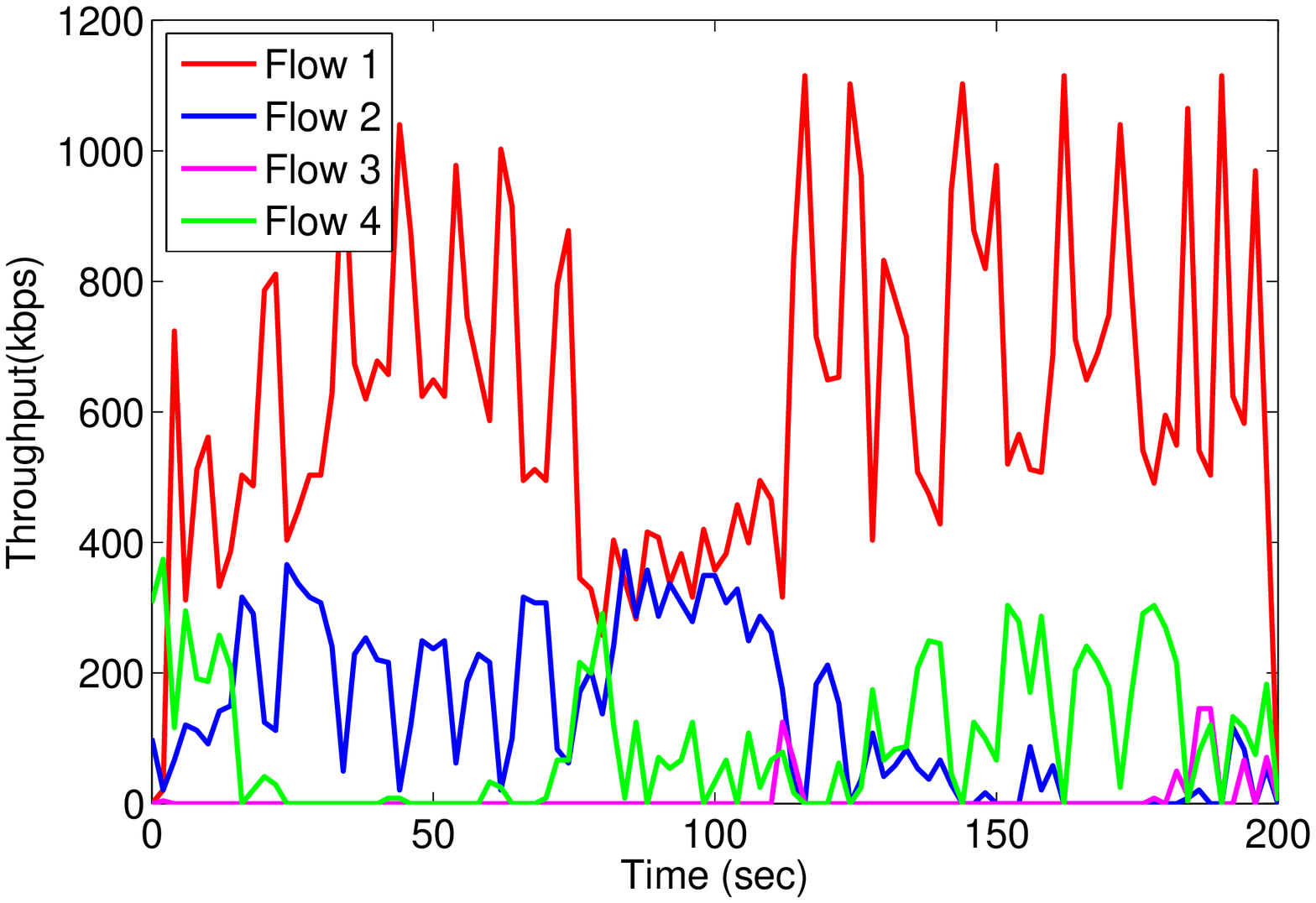}}} \hspace{-0pt}
\end{center}
\begin{center}
\vspace{-15pt}
\caption{\label{fig:grid_thrpt_time_results} \scriptsize Throughput vs. time in the grid topology for TCP-SACK and TCP-Vegas. There are four flows and the links are not lossy.
}
\end{center}
\vspace{-15pt}
\end{figure}

\begin{figure}[t!]
\vspace{-0pt}
\begin{center}
\subfigure[\scriptsize Throughput]{{\includegraphics[width=4cm]{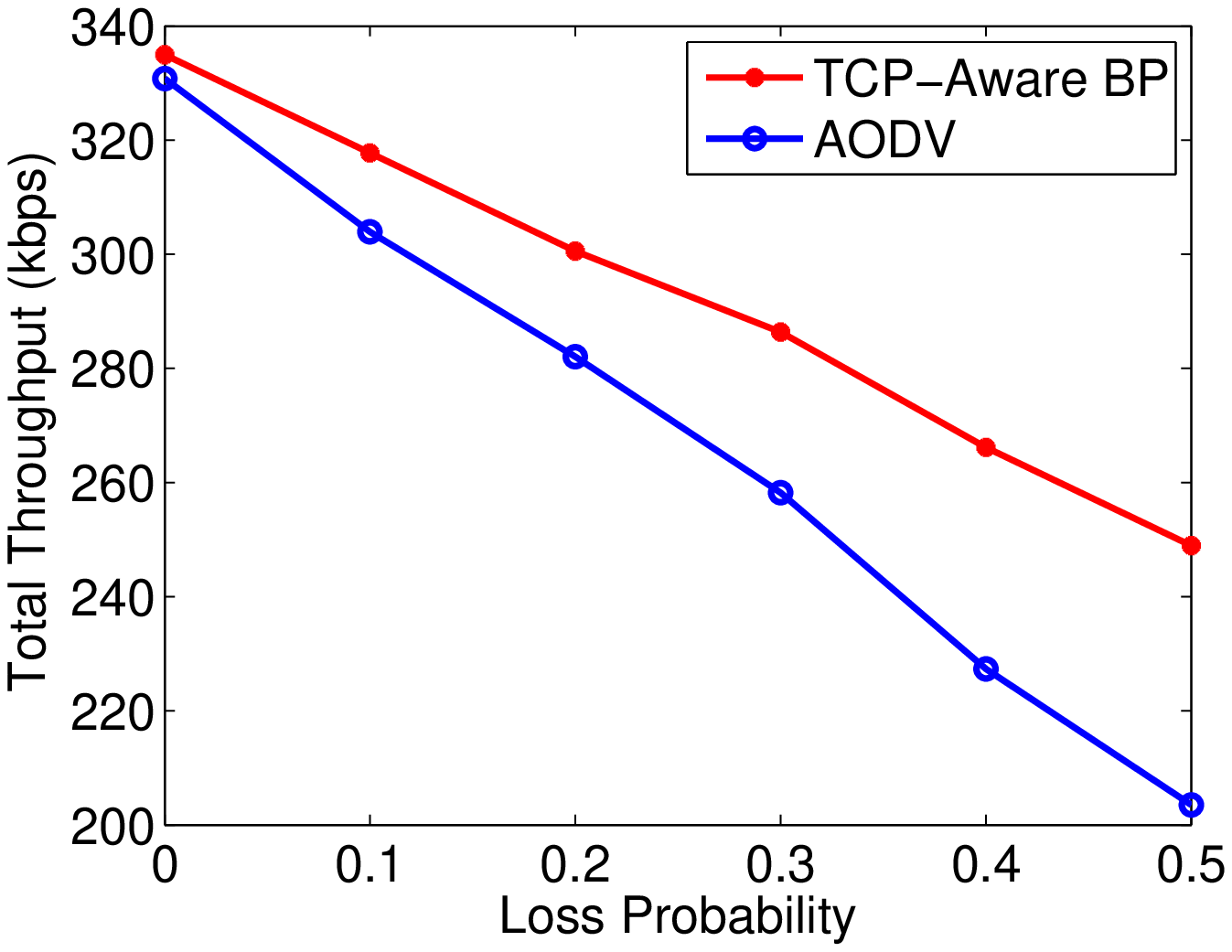}}}
\subfigure[\scriptsize Fairness]{{\includegraphics[width=4cm]{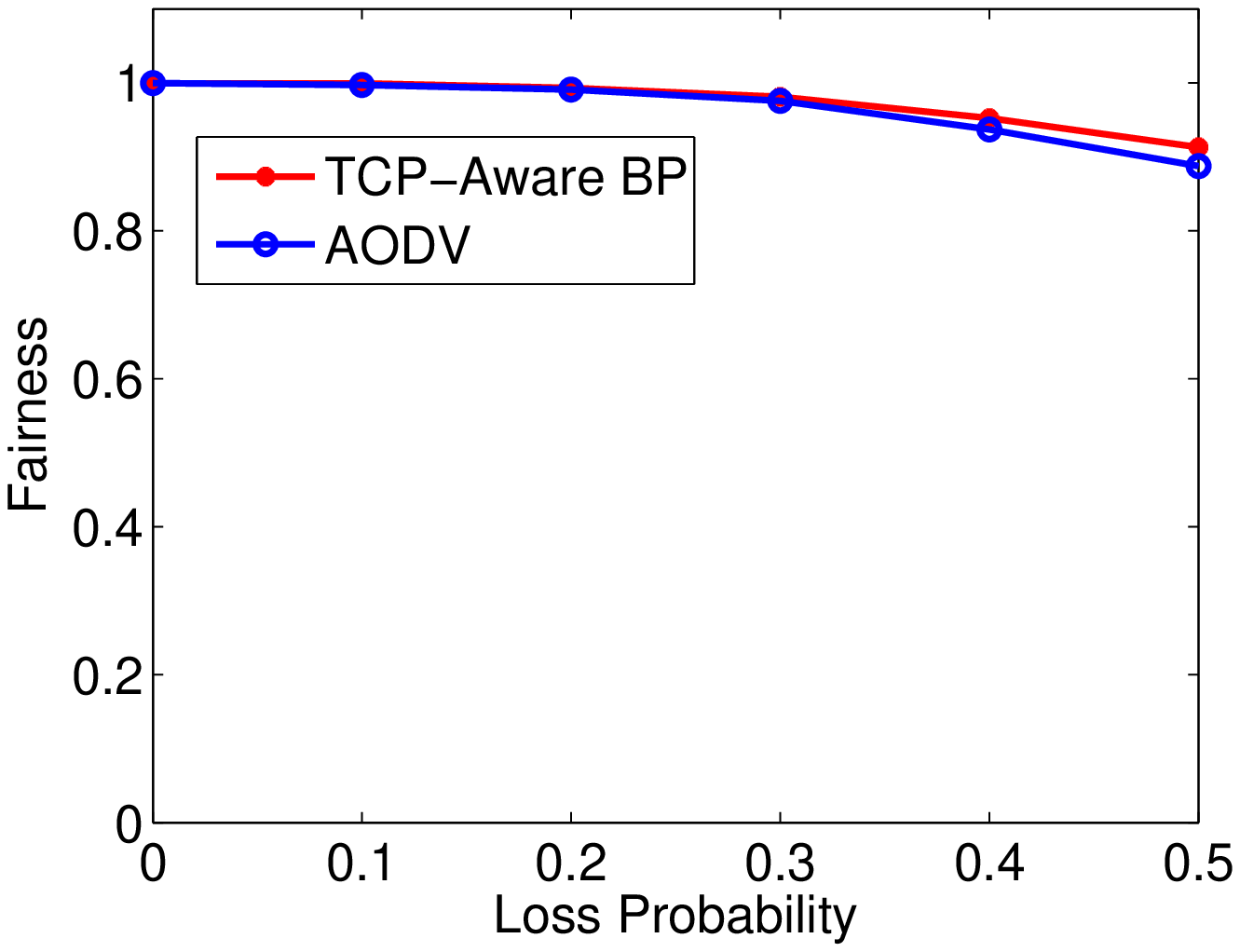}}}
\end{center}
\begin{center}
\vspace{-5pt}
\caption{\label{fig:grid_thrpt_vs_loss_sack} \scriptsize Throughput and fairness vs. average packet loss rate for TCP-aware BP and AODV in the grid topology. There are four TCP flows transmitted from the gateway to four distinct nodes. Half of the links are lossy. The version of TCP is TCP-SACK.
}
\end{center}
\vspace{-15pt}
\end{figure}

\begin{figure}[t!]
\vspace{-0pt}
\begin{center}
\subfigure[\scriptsize Throughput]{{\includegraphics[width=4cm]{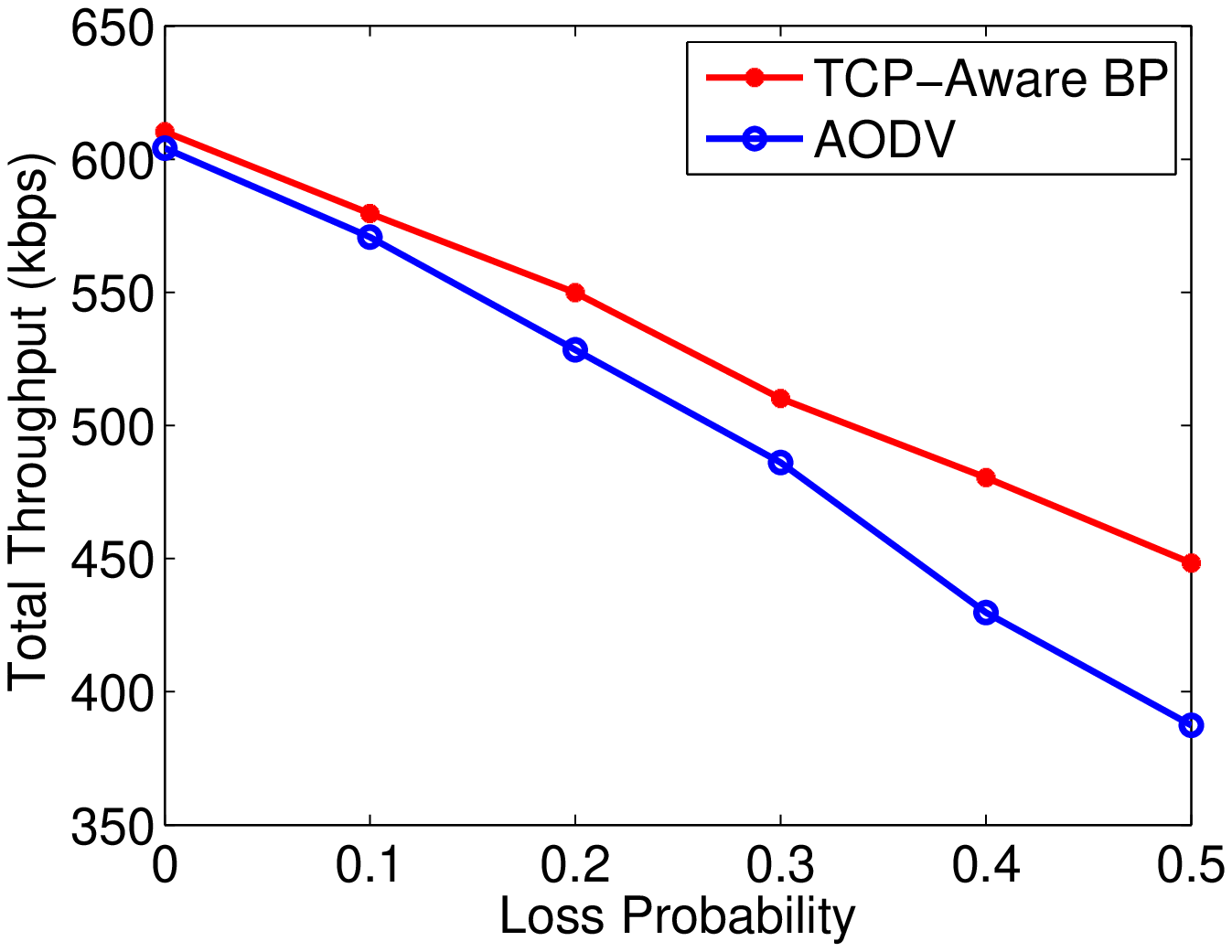}}}
\subfigure[\scriptsize Fairness]{{\includegraphics[width=4cm]{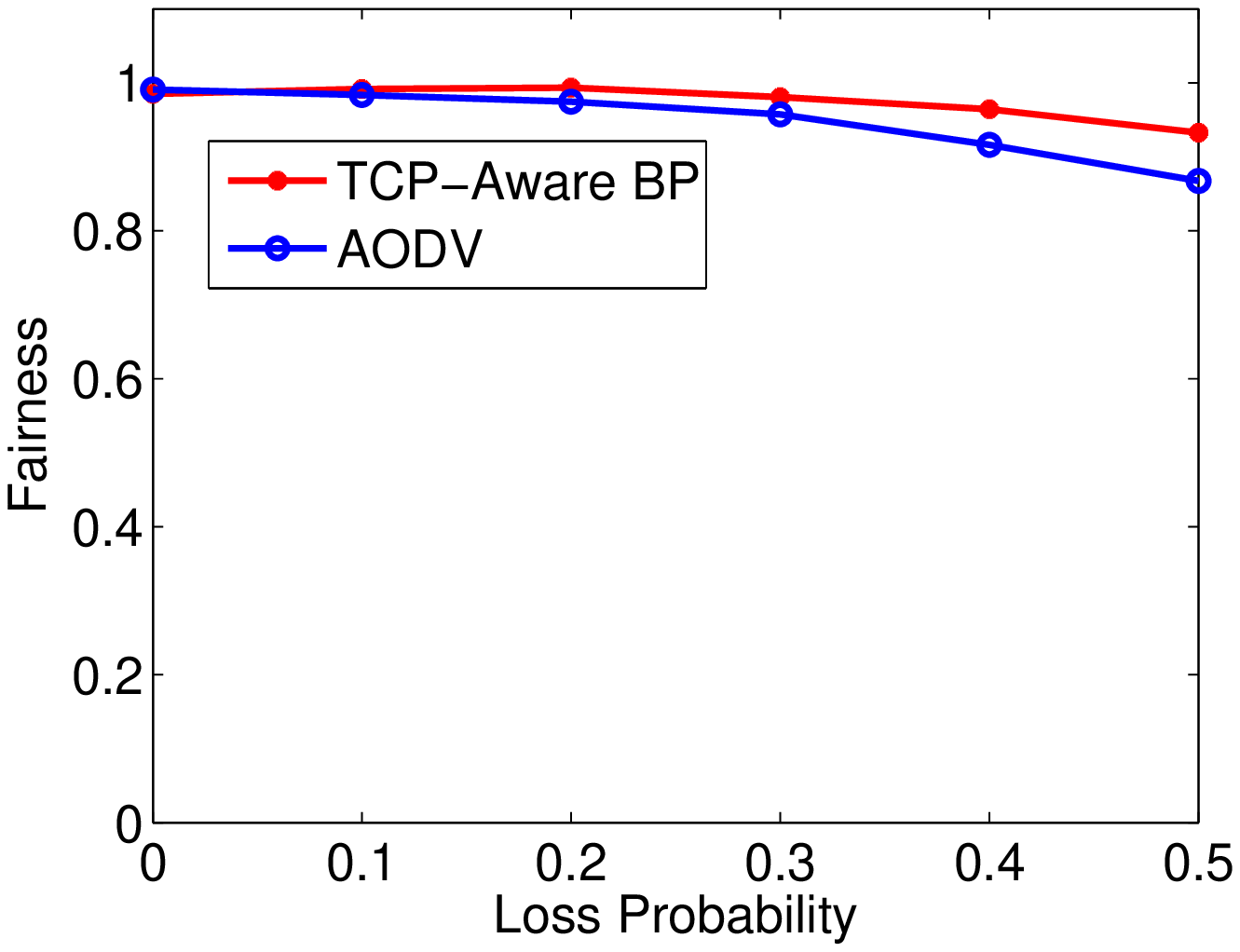}}}
\end{center}
\begin{center}
\vspace{-5pt}
\caption{\label{fig:grid_thrpt_vs_loss_vegas} \scriptsize Throughput and fairness vs. average packet loss rate for TCP-aware BP and AODV in the grid topology. There are four TCP flows transmitted from the gateway to four distinct nodes. Half of the links are lossy. The version of TCP is TCP-Vegas.}
\end{center}
\vspace{-15pt}
\end{figure}

Let us consider the grid topology shown in Fig.~\ref{fig:topologies}. Four flows are transmitted from the gateway to four distinct nodes, which are randomly chosen. Half of the links, chosen at random, are lossy with loss probability ranging between $0-0.5$. Fig.~\ref{fig:grid_thrpt_time_results} shows throughput vs. time graphs for TCP-aware BP and classical BP. It is seen that all four flows could survive in TCP-aware BP for both TCP-SACK and TCP-Vegas, while one or more flows do not survive in classical BP. Fig.~\ref{fig:grid_thrpt_vs_loss_sack} shows throughput and fairness vs. average loss probability results for TCP-aware BP and AODV for TCP-SACK. TCP-aware BP improves throughput significantly as compared to AODV without violating fairness. 
Fig.~\ref{fig:grid_thrpt_vs_loss_vegas} shows that TCP-aware BP improves throughput significantly as compared to AODV when TCP-Vegas is employed. This shows the effectiveness of our scheme in delay-based TCP versions.


\begin{figure}[t!]
\vspace{-0pt}
\begin{center}
\subfigure[\scriptsize TCP-Aware BP]{{\includegraphics[width=2.7cm]{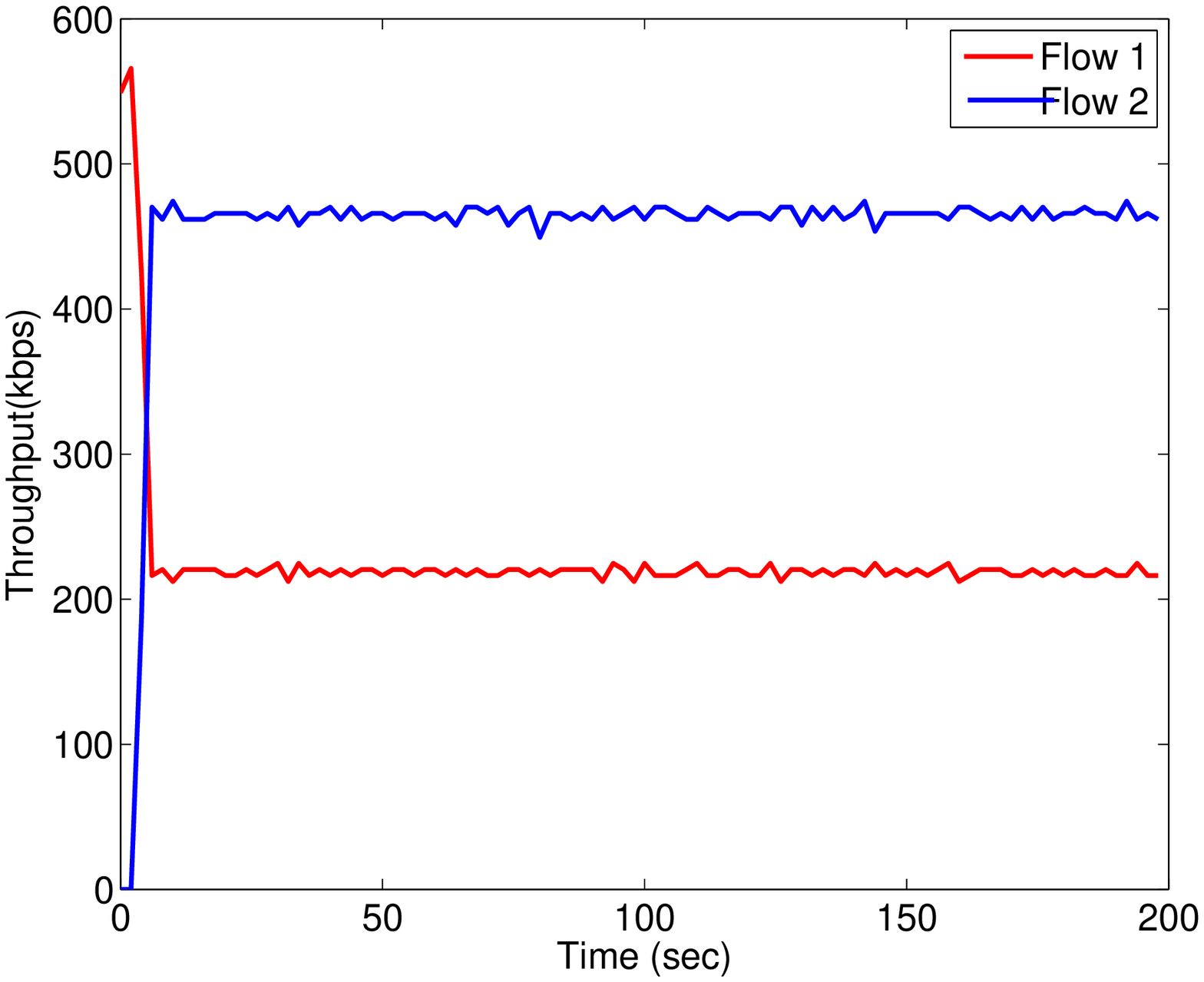}}}
\subfigure[\scriptsize Classical BP]{{\includegraphics[width=2.8cm]{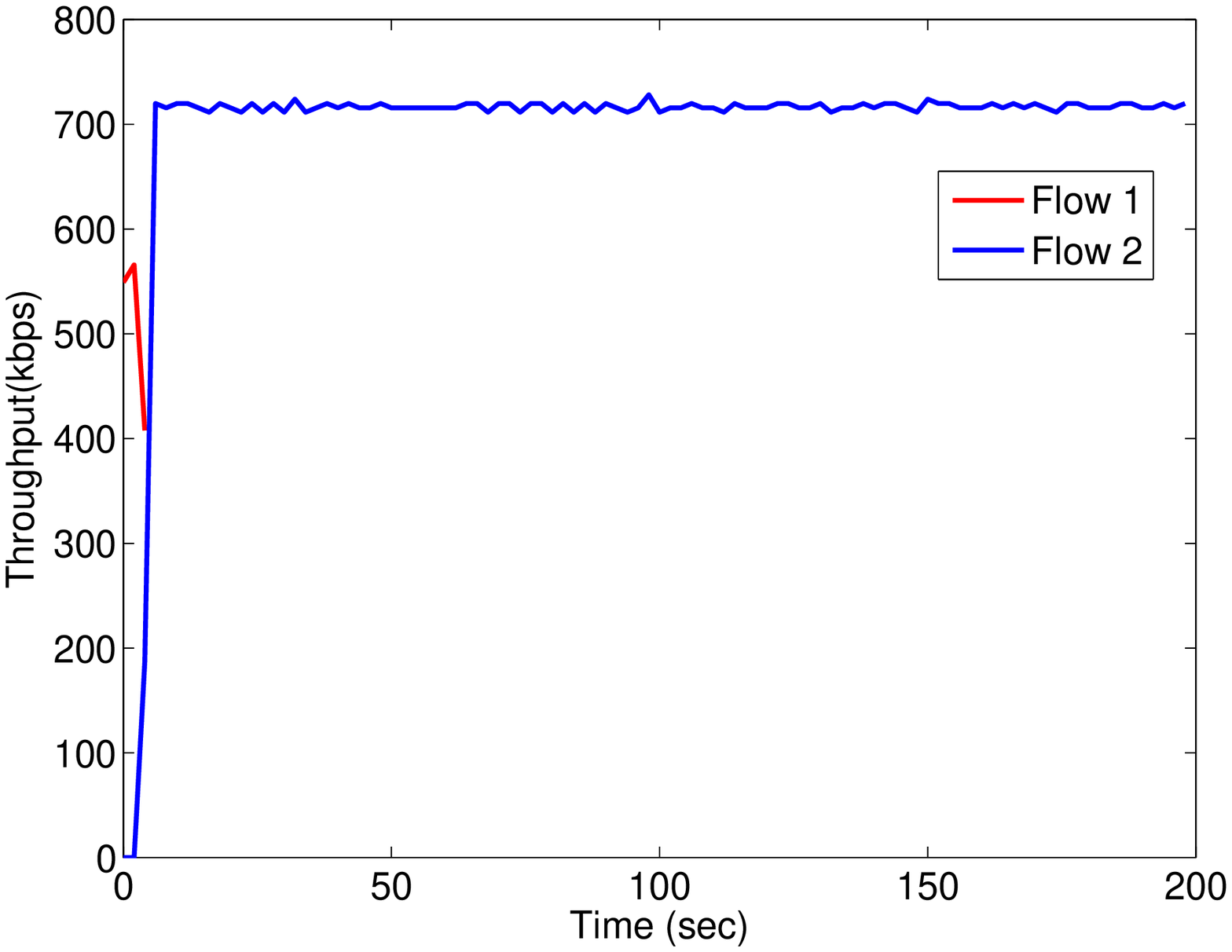}}} \hspace{-0pt}
\subfigure[\scriptsize AODV]{{\includegraphics[width=2.8cm]{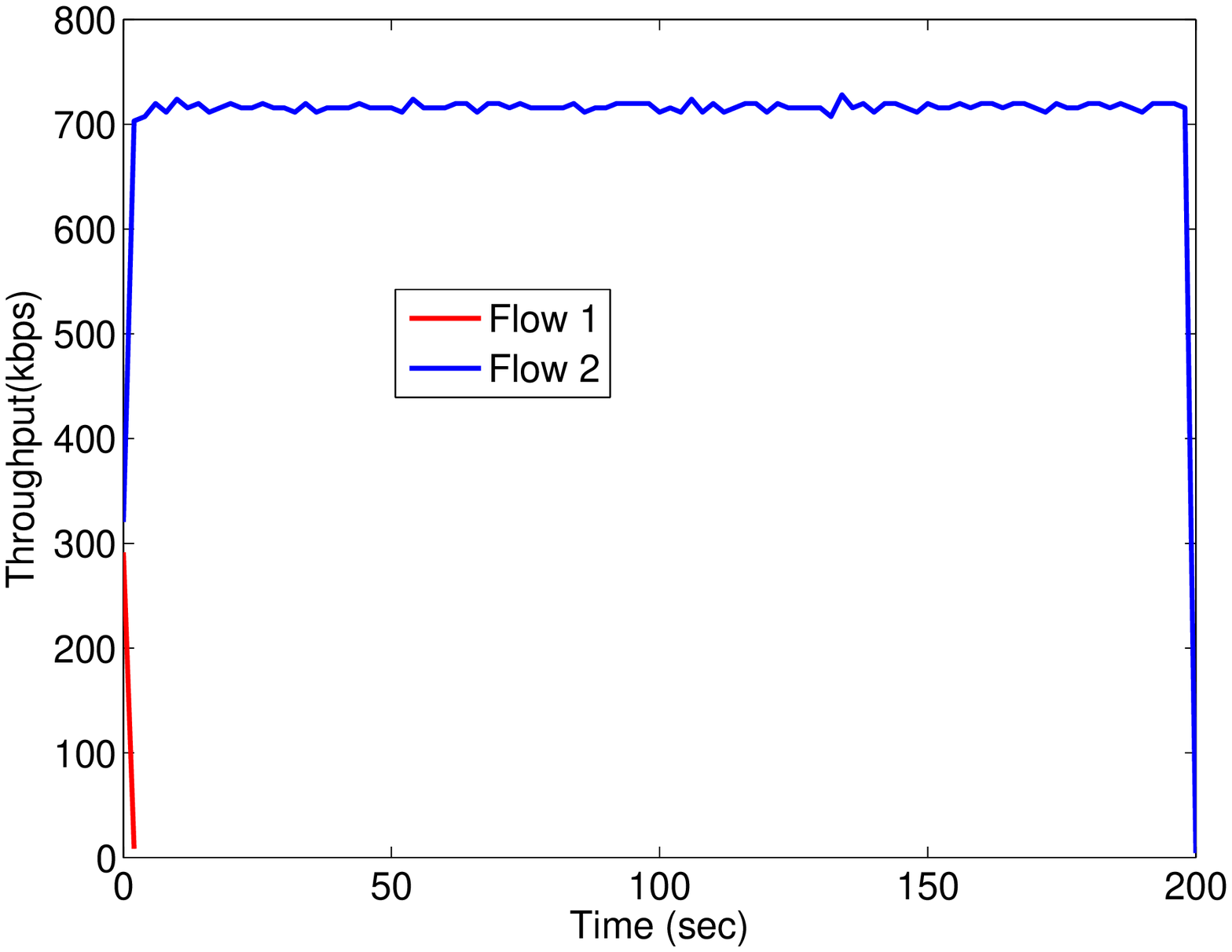}}}
\end{center}
\begin{center}
\vspace{-5pt}
\caption{\label{fig:diamond_thrpt_time_results_tcp_udp} \scriptsize Throughput vs. time in the diamond topology for TCP-SACK. There are two flows; Flow 1 is a TCP flow, transmitted from node $A$ to node $B$, and Flow 2 is a non-TCP flow, transmitted from node $A$ to node $D$. The links are not lossy.
}
\end{center}
\vspace{-15pt}
\end{figure}

\begin{figure}[t!]
\vspace{-0pt}
\begin{center}
\subfigure[\scriptsize Throughput]{{\includegraphics[width=4cm]{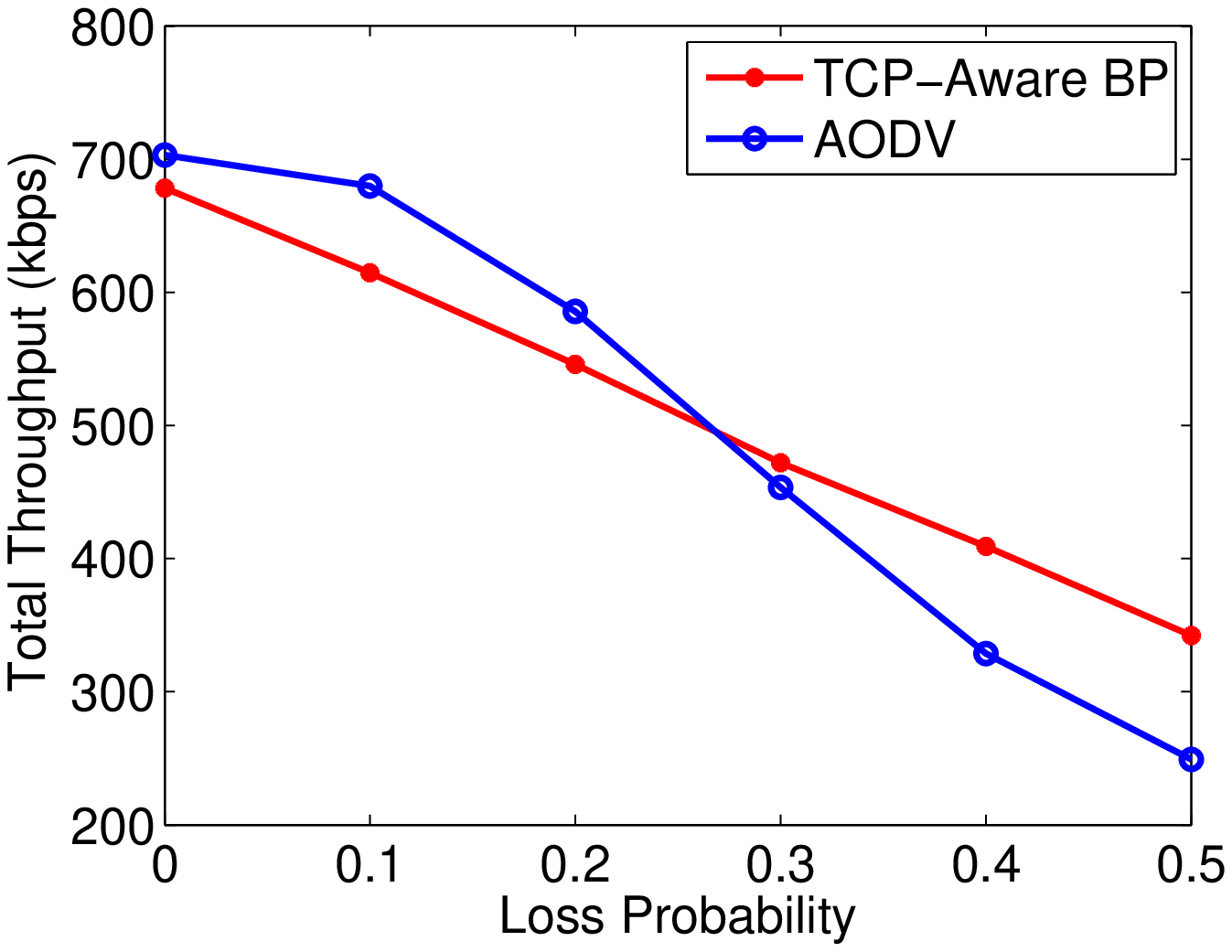}}}
\subfigure[\scriptsize Fairness]{{\includegraphics[width=4cm]{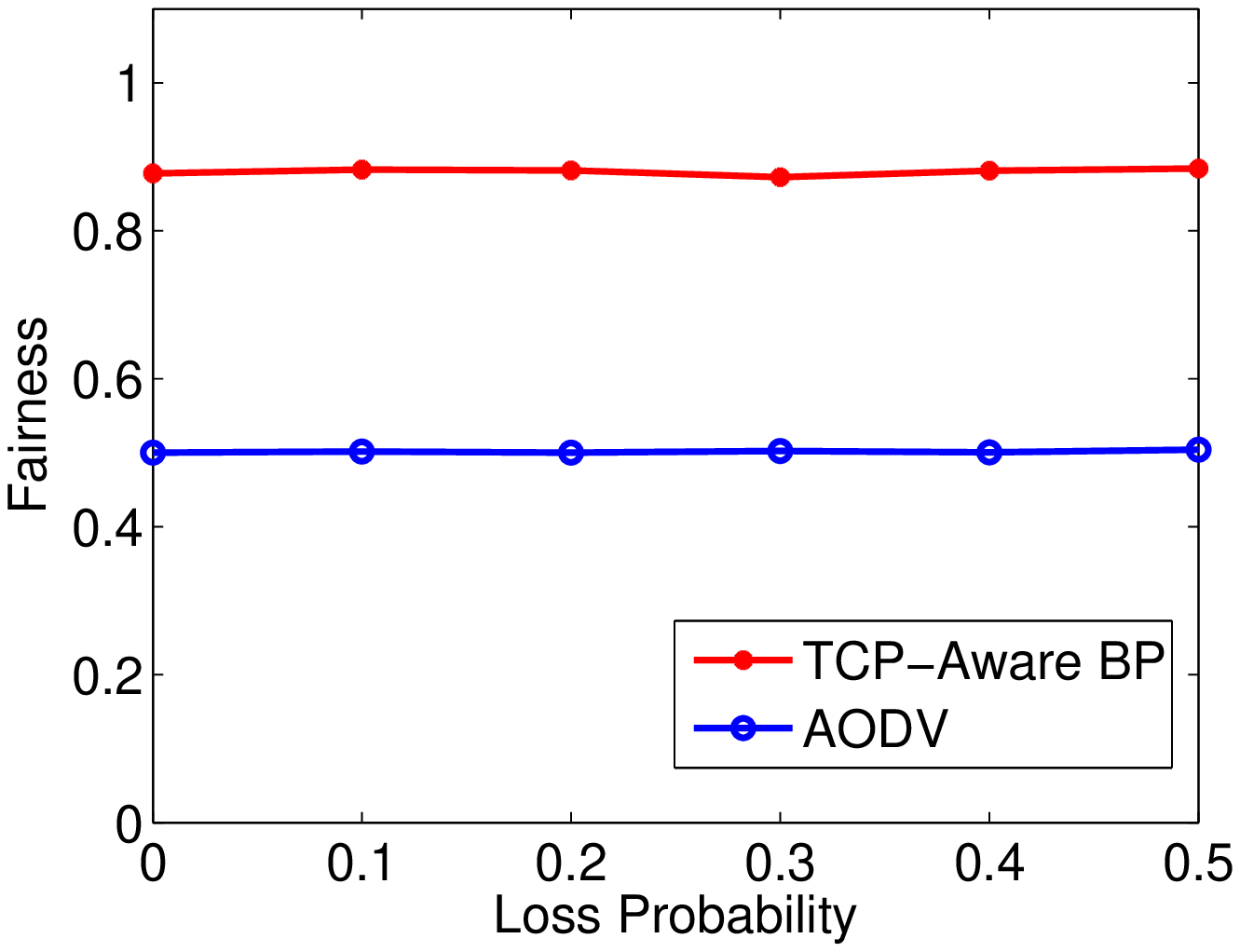}}}
\end{center}
\begin{center}
\vspace{-5pt}
\caption{\label{fig:diamond_thrpt_vs_loss_sack_UDP_TCP} \scriptsize Throughput and fairness vs. average packet loss rate for TCP-aware BP and AODV in the diamond topology. There are two flows transmitted from node $A$ to $B$ (Flow 1, \ie TCP flow) and $A$ to $D$ (Flow 2, \ie non-TCP flow). The link $A-B$ is a lossy link. The version of TCP is TCP-SACK.
}
\end{center}
\vspace{-25pt}
\end{figure}

As mentioned in Section~\ref{sec:opt}, there may be both TCP and non-TCP flows in the system, and non-TCP flows should be controlled in a TCP-friendly manner so that TCP flows could survive when non-TCP flows are on. Therefore, a flow control algorithm is presented in Eq.~(\ref{eq:flow_control}) for non-TCP flows. Now, we evaluate this scenario in the diamond topology with two flows. Flow 1 is a TCP flow (TCP-SACK) transmitted from node $A$ to node $B$, and Flow 2 is a non-TCP flow transmitted from node $A$ to node $D$. In our TCP-aware BP framework, the non-TCP flow is regulated by Eq.~(\ref{eq:flow_control}). The parameters in Eq.~(\ref{eq:flow_control}) are set as; $M=50$, $g(x_s(t))=log(x_s(t))$, $\forall t, s \in \Sset$. The implementation details including TCP-friendly parameter selection are provided in \cite{tcp_aware_bp_tech_rep}.
Fig.~\ref{fig:diamond_thrpt_time_results_tcp_udp} shows throughput vs. time graph of TCP-aware BP, classical BP, and AODV. The TCP flow does not survive in classical BP as packets are trapped in the buffers. It does not survive with AODV as well, because uncontrolled non-TCP flows (\ie UDP flows) occupy buffers and TCP packets are constantly dropped from the buffers, which reduces TCP throughput. Yet, both TCP and non-TCP flows survive together in in TCP-aware BP thanks to TCP-aware routing and scheduling, and TCP-friendly flow control for non-TCP flows. Fig.~\ref{fig:diamond_thrpt_vs_loss_sack_UDP_TCP} shows the throughput improvement performance of TCP-aware BP as compared to AODV in the same setup for different packet loss probabilities. At low loss probabilities, although the throughput of AODV is better than TCP-aware BP, the fairness graph (and Fig.~\ref{fig:diamond_thrpt_time_results_tcp_udp} for no-loss) shows that the fairness of AODV is very low, which means that the TCP flow does not survive. At higher loss probabilities, TCP-aware BP is better than AODV thanks to choosing better routes and schedules as compared to AODV.

\section{Related Work}\label{sec:related}
Backpressure, a routing and scheduling framework over communication networks \cite{tass_eph1}, \cite{tass_eph2} has generated a lot of research interest \cite{neely_book}, mainly in wireless ad-hoc networks. 
It has also been shown that backpressure can be combined with flow control to provide utility-optimal operation guarantee \cite{neely_mod}, \cite{stolyar_greedy}.

The strengths of backpressure have recently increased the interest on practical implementation of backpressure over wireless networks. Backpressure has been implemented over sensor networks \cite{routing_wtht_routes} and wireless multi-hop networks \cite{xpress}. The multi-receiver diversity has been explored in wireless networks using backpressure in \cite{javidi_diversity}. The 802.11 compliant version of enhanced backpressure is evaluated in \cite{choumas}. Backpressure routing and rate control for intermittently connected networks was developed in \cite{backpressure_for_icns}. 

Backpressure routing and (max-weight) scheduling with TCP over wireless has been considered in the literature. At the link layer, \cite{DiffQ}, \cite{umut_stolyar}, propose, analyze, and evaluate link layer backpressure-based implementations with queue prioritization and congestion window size adjustment. The interaction of TCP with backpressure in \cite{DiffQ} and \cite{umut_stolyar} is handled by updating the TCP congestion window evolution mechanism. In particular, if the queue size (at the TCP source) increases, the window size is reduced, otherwise, the window size is increased. Multi-path TCP scheme is implemented over wireless mesh networks \cite{horizon} for routing and scheduling packets using a backpressure based heuristic, which avoids incompatibility with TCP. 
Max-weight scheduling is updated in \cite{ghaderi_tcp_theoric} to make decisions based only on MAC level queue size information. Although \cite{ghaderi_tcp_theoric} considers window based flow control mechanism similar to TCP, it does not consider existing TCP flavors. 
The main differences in our work are: (i) we consider the incompatibility of TCP with backpressure, and develop TCP-aware backpressure framework to address the incompatibilities, (ii) TCP-aware backpressure provides the same stability and utility-optimal operation guarantees as classical backpressure, (iii) we do not make any changes at the TCP source, (iv) we employ network coding to gracefully combine TCP and TCP-aware backpressure.

Maximum weight matching (MWM) is a switch scheduling algorithm and has similar properties as the max-weight scheduling algorithm and backpressure. Similar to the backpressure, there is incompatibility between TCP and MWM \cite{TCP_MWM_switch1}, \cite{TCP_MWM_switch2}. Yet, we consider backpressure routing and scheduling over wireless networks rather than switch scheduling, and we take a holistic approach to address this problem; \ie we propose TCP-aware backpressure to make TCP and backpressure compatible.

\section{Conclusion}\label{sec:conclusion}
We proposed TCP-aware backpressure routing and scheduling to address the incompatibility of TCP and backpressure while exploiting the performance of backpressure routing and scheduling over wireless networks. TCP-aware backpressure is developed by taking into account the behavior of TCP flows, and gracefully combines TCP and backpressure without making any changes to the TCP protocol. Simulations in ns-2 demonstrate that TCP-aware backpressure improves throughput of TCP flows significantly and provides fairness across competing TCP flows.



\bibliographystyle{IEEEtran}

\begin{thebibliography}{}
\bibitem{tass_eph1} L. Tassiulas, A. Ephremides, ``Stability properties of constrained queueing systems and scheduling policies for maximum throughput in multihop radio networks,'' {\em in IEEE Trans. on Auto. Control}, vol.~37(12), Dec. 1992.

\bibitem{tass_eph2} L. Tassiulas, A. Ephremides, ``Dynamic server allocation to parallel queues with randomly varying connectivity,'' {\em in IEEE ToIT}, vol.~39(2), March 1993.

\bibitem{neely_mod} M. J. Neely, E. Modiano, C. Li, ``Fairness and optimal stochastic control for heterogeneous networks,'' {\em in IEEE/ACM ToN}, vol.~16(2), April 2008.

\bibitem{tcp_compound} K.~Tan,J.~Song, Q.~Zhang, M.~Sridharan, ``A compound TCP approach for high-speed and long distance networks,'' {\em in Proc. of IEEE INFOCOM}, Barcelona, Spain, April 2006.

\bibitem{tcp_cubic} S.~Ha, I.~Rhee, L.~Xu, ``CUBIC: a new TCP-friendly high-speed TCP variant,'' {\em in SIGOPS Oper. Syst. Rev.}, vol.~42(5), July 2008.

\bibitem{ns2} The Network Simulator - ns-2, Version 2.35, {\em available at www.isi.edu/nsnam/ns/}.

\bibitem{neely_book} M. J. Neely, ``Stochastic network optimization with application to communication and queueing systems,'' Morgan \& Claypool, 2010.


\bibitem{tcp_aware_bp_tech_rep} H.~Seferoglu, E.~Modiano, ``TCP-aware backpressure routing and scheduling,'' Tech. Report, available at {\em newport.eecs.uci.edu/{\texttildelow}hseferog/, http://www.mit.edu/{\texttildelow}hseferog/}.

\bibitem{tutorial_doyle} M. Chiang, S. T. Low, A. R. Calderbank, J. C. Doyle, ``Layering as optimization decomposition: a mathematical theory of network architectures,'' \emph{in Proceedings of the IEEE}, vol. 95(1), Jan. 2007.

\bibitem{lin_schroff_tutorial} X. Lin, N. B. Schroff, R. Srikant, ``A tutorial on cross-layer optimization in wireless networks,'' {\em in IEEE JSAC}, vol. 24(8), Aug. 2006.

\bibitem{diffmax} H.~Seferoglu, E.~Modiano, ``Diff-Max: separation of routing and scheduling in backpressure-based wireless Networks,'' {\em in Proc. of IEEE INFOFOCM}, Turin, Italy, April, 2013.

\bibitem{locbui} L.~X.~Bui, R.~Srikant, A.~Stolyar, ``A novel architecture for reduction of delay and queueing structure complexity in the back-pressure algorithm,'' {\em in IEEE/ACM Transactions on Networking}, vol.~19(6), Dec. 2011.

\bibitem{NC_meets_TCP} J. K. Sundararajan, D. Shah, M. Medard, M. Mitzenmacher, J. Barros, ``Network coding meets TCP,'' {\em in Proc. of IEEE INFOCOM}, Rio de Janeiro, Brazil, April 2009.

\bibitem{multipath_tcp_toledo} 	S. Gheorghiu, A. L. Toledo, P. Rodriguez, ``Multi-path TCP with network coding for wireless mesh networks,'' {\em in Proc. of IEEE ICC}, Cape Town, South Africa, May 2010.

\bibitem{i2nc} H. Seferoglu, A. Markopoulou, K. K. Ramakrishnan, ``I$^2$NC: intra- and inter-session network coding for unicast flows in wireless networks,'' {\em in Proc. of IEEE INFOCOM}, Shanghai, China, April 2011.

\bibitem{practical_NC} P. A. Chou, Y. Wu,``Network coding for the Internet and wireless networks,'' {\em in IEEE Signal Proc. Magazine}, vol. 24(5), Sept. 2007.

\bibitem{twsly_tcp} J.~Padhye, V.~Firoiu, D.~Towsley, J.~Kurose, ``Modeling TCP throughput: a simple model and its empirical validation,'' {\em in Proc. of ACM SIGCOMM}, Vancouver, Canada, Sep. 1998.

\bibitem{low_tcp} S.~Low, ``A duality model of TCP and queue management algorithms,'' {\em in IEEE/ACM Transactions on Networking}, vol.~11(4), Aug. 2003.

\bibitem{aodv} C. Perkins, E. Belding-Royer, S. Das, ``Ad hoc on-demand distance vector (AODV) routing,'' {\em RFC 3561, IETF}, July 2003.

\bibitem{fairness_index} R. K. Jain, ``The art of computer systems performance analysis: techniques for experimental design, measurement, simulation, and modeling,'' John Wiley \& Sons, April 1991.

\bibitem{stolyar_greedy} A. L. Stolyar, ``Greedy primal dual algorithm for dynamic resource allocation in complex networks,'' {\em in Queuing Systems}, vol. 54, 2006.

\bibitem{routing_wtht_routes} S. Moeller, A. Sridharan, B. Krishnamachari, O. Gnawali, ``Routing without routes: the backpressure collection protocol,'' {\em in Proc. of ACM IPSN}, Stockholm, Sweden, April 2010.

\bibitem{xpress} R. Laufer, T. Salonidis, H. Lundgren, P. L. Guyadec, ``XPRESS: a cross-layer backpressure architecture for wireless multi-hop networks,'' {\em in Proc. of ACM MobiCom}, Las Vegas, NV, Sep. 2011.

\bibitem{javidi_diversity} A. A. Bhorkar, T. Javidi, A. C. Snoereny, ``Achieving congestion diversity in wireless ad-hoc networks,'' {\em in Proc. of IEEE INFOCOM}, Shanghai, China, April 2011.

\bibitem{choumas} K. Choumas, T. Korakis, I. Koutsopoulos, L. Tassiulas, ``Implementation and end-to-end throughput evaluation of an IEEE 802.11 compliant version of the enhanced-backpressure algorithm,'' {\em in Proc. of TridentCom},  Thessaloniki, Greece, June 2012.

\bibitem{backpressure_for_icns} J. Ryu, V. Bhargava, N. Paine, S. Shakkottai, ``Backpressure routing and rate control for ICNs,'' {\em in Proc. of ACM MobiCom}, Chicago, IL, Sep. 2010.

\bibitem{DiffQ} A. Warrier, S. Janakiraman, S. Ha, I. Rhee, ``DiffQ: practical differential backlog congestion control for wireless networks,'' {\em in Proc. of IEEE INFOCOM}, Rio de Janerio, Brazil, April 2009.

\bibitem{umut_stolyar} U. Akyol, M. Andrews, P. Gupta, J. Hobby, I. Saniee, A. Stolyar, ``Joint scheduling and congestion control in mobile ad-hoc networks,'' {\em in Proc. of IEEE INFOCOM}, Phoenix, AZ, April 2008.

\bibitem{horizon} B. Radunovic, C. Gkantsidis, D. Gunawardena, P. Key, ``Horizon: balancing TCP over multiple paths in wireless mesh network,'' {\em in Proc. of ACM MobiCom}, San Francisco, CA, Sep. 2008.

\bibitem{ghaderi_tcp_theoric} J. Ghaderi, T. Ji, R. Srikant, ``Connection-level scheduling in wireless networks using only MAC-layer information,'' {\em in Proc. of IEEE INFOCOM}, Orlando, FL, March 2012.

\bibitem{TCP_MWM_switch1} A. Shpiner, I. Keslassy, ``Modeling the interactions of congestion control and switch scheduling,'' {\em in Computer Networks}, vol. 55(6), April 2011.

\bibitem{TCP_MWM_switch2} P. Giaccone, E. Leonardi, F. Neri, ``On the interaction between TCP-like sources and throughput-efficient scheduling policies,'' {\em in Elsevier}, 2013.

\bibitem{gamma_ksi_approximization} Y. Yi, A. Prouti\`{e}re, and M. Chiang, ``Complexity in wireless scheduling: impact and tradeoffs,'' {\em in Proc. of ACM MobiHoc}, Hong Kong, China, May 2008.


\end{thebibliography}

\section*{Appendix A: Proof of Theorem~\ref{theorem1}}
The proof of Theorem~\ref{theorem1} directly follows from $(\gamma, \xi)$ (or $C$-) approximation in \cite{gamma_ksi_approximization}, \cite{neely_book}. We provide the proof in this section for completeness. Let $\oset{*}{f}_{i,j}^{s}(t)$ be the optimal decision when $D_{i,j}^{s}(t) = U_{i}^{s}(t) - U_{j}^{s}(t)$ in Eq.~(\ref{eq:per_flow_difference}) (note that this is the classical backpressure), while $f_{i,j}^{s}(t)$ be the decision when $D_{i,j}^{s}(t) = \max\{K,U_{i}^{s}(t)\} - U_{j}^{s}(t)$. If the following inequality holds, the policy that makes decision based on $D_{i,j}^{s}(t) = \max\{K,U_{i}^{s}(t)\} - U_{j}^{s}(t)$ stabilizes the queues.
\begin{align} \label{eq:app_1}
& \sum_{(i,j) \in \Lset} \sum_{s \in \Sset} f_{i,j}^{s}(t) (U_{i}^{s}(t) - U_{j}^{s}(t)) \geq \sum_{(i,j) \in \Lset} \sum_{s \in \Sset} \oset{*}{f}_{i,j}^{s}(t) (U_{i}^{s}(t) \nonumber \\
& - U_{j}^{s}(t)) - Z,
\end{align} where $Z$ is a finite constant. Let us first show that Eq.~(\ref{eq:app_1}) holds. Consider the inequality; $\sum_{(i,j) \in \Lset} \sum_{s \in \Sset} (U_{i}^{s}(t) - U_{j}^{s}(t)) \oset{*}{f}_{i,j}^{s}(t)  \leq \sum_{(i,j) \in \Lset} \sum_{s \in \Sset} \max \{K, U_{i}^{s}(t)\} \oset{*}{f}_{i,j}^{s}(t)$. This inequality holds, because $\max \{K, U_{i}^{s}(t) \}$ $\geq$ $U_{i}^{s}(t) $. Also, considering that $f_{i,j}^{s}(t)$ is the optimal decision when the backlog difference is $\max \{K,U_{i}^{s}(t)\} - U_{j}^{s}(t)$, the term $(\max \{K,$ $U_{i}^{s}(t)\}$ $- U_{j}^{s}(t)) f_{i,j}^{s}(t)$ should be greater than $(\max \{K,U_{i}^{s}(t)\}$ $- U_{j}^{s}(t))$ $\oset{*}{f}_{i,j}^{s}(t)$. Therefore, the inequality is expressed as; $ \sum_{(i,j) \in \Lset} \sum_{s \in \Sset} (U_{i}^{s}(t) - U_{j}^{s}(t)) \oset{*}{f}_{i,j}^{s}(t)  \leq \sum_{(i,j) \in \Lset} \sum_{s \in \Sset} \max$ $\{K, U_{i}^{s}(t)\} f_{i,j}^{s}(t)$. By adding and removing terms, and noting that $\max \{K,U_{i}^{s}(t)\} - U_{i}^{s}(t) \leq K$ and $f_{i,j}^{s}(t) \leq F_{\max}$ such that $F_{\max} \geq R_{i,j}$, the following holds;
\begin{align} \label{eq:app_7}
& \sum_{(i,j) \in \Lset} \sum_{s \in \Sset} (U_{i}^{s}(t) - U_{j}^{s}(t)) f_{i,j}^{s}(t)  \geq \sum_{(i,j) \in \Lset} \sum_{s \in \Sset} (U_{i}^{s}(t) \nonumber \\
& - U_{j}^{s}(t)) \oset{*}{f}_{i,j}^{s}(t) - |\Lset| |\Sset| K F_{\max}
\end{align}  Eq.~(\ref{eq:app_7}) verifies that Eq.~(\ref{eq:app_1}) holds considering that $Z = $ $|\Lset|$ $|\Sset|$ $K$ $F_{\max}$.
Note that Eq.~(\ref{eq:app_7}) is equivalent to;
\begin{align} \label{eq:app_8}
& \sum_{i \in \Nset} \sum_{s \in \Sset} U_{i}^{s}(t) (\sum_{j \in \Nset} f_{i,j}^{s}(t) - \sum_{j \in \Nset} f_{j,i}^{s}(t))  \geq \sum_{i \in \Nset} \sum_{s \in \Sset} U_{i}^{s}(t) \nonumber \\
& (\sum_{j \in \Nset} \oset{*}{f}_{i,j}^{s}(t) - \sum_{j \in \Nset} \oset{*}{f}_{j,i}^{s}(t)) - |\Lset| |\Sset| K F_{\max}
\end{align}
Now, let us define the Lyapunov function as; $L(\boldsymbol U(t)) =  \sum_{i \in \Nset} \sum_{s \in \Sset} U_{i}^{s}(t)^{2}$, where $\boldsymbol U(t) = \{ U_{i}^{s}(t) \}_{i \in \Nset, s \in \Sset}$ and $U_{i}^{s}(t)$ evolves according to Eq.~(\ref{eq:queue_U}). Let the Lyapunov drift be $\Delta (\boldsymbol U(t)) =  E[ L(\boldsymbol U(t+1)) - L(\boldsymbol U(t)) | \boldsymbol U(t) ]$, which is equal to $\Delta (\boldsymbol U(t)) \leq  E[ \sum_{i \in \Nset} \sum_{s \in \Sset} (U_{i}^{s}(t+1))^{2} - \sum_{i \in \Nset} \sum_{s \in \Sset}$ $(U_{i}^{s}(t))^{2} | \boldsymbol U(t) ]$.
Using the fact that $(\max(Q-b,0)+A)^2 \leq Q^2 + A^2 + b^2 + 2Q(A-b)$, we have; $\Delta (\boldsymbol U(t)) \leq E[  \sum_{i \in \Nset} \sum_{s \in \Sset}$ $(\sum_{j \in \Nset} f_{i,j}^{s}(t))^{2} + \sum_{i \in \Nset} \sum_{s \in \Sset} (\sum_{j \in \Nset} f_{i,j}^{s}(t) + A_{s}(t)1_{[i=o(s)]})^{2}$ $+ \sum_{i \in \Nset} \sum_{s \in \Sset} 2U_{i}^{s}(t) ( \sum_{j \in \Nset} f_{j,i}^{s}(t) - \sum_{j \in \Nset} f_{i,j}^{s}(t) ) + \sum_{i \in \Nset}$ $\sum_{s \in \Sset} 2U_{i}^{s}(t)A_{s}(t)1_{[i=o(s)]} | \boldsymbol U(t) ]$. Noting that there always exist a finite constant $B$ such that $B \geq E[ \sum_{i \in \Nset} \sum_{s \in \Sset} (\sum_{j \in \Nset}$ $f_{i,j}^{s}(t))^{2} + \sum_{i \in \Nset} \sum_{s \in \Sset} (\sum_{j \in \Nset}$ $f_{i,j}^{s}(t)+ A_{s}(t)1_{[i=o(s)]})^{2} | \boldsymbol U(t)]$, we have;
\begin{align} \label{eq:lyap_6}
\Delta (\boldsymbol U(t)) & \leq B - 2E[ \sum_{i \in \Nset} \sum_{j \in \Nset} \sum_{s \in \Sset} U_{i}^{s}(t)(f_{i,j}^{s}(t) - f_{j,i}^{s}(t)) \nonumber \\
& + 2 \sum_{i \in \Nset} \sum_{s \in \Sset} U_{i}^{s}(t) A_{s}(t)1_{[i=o(s)]} | \boldsymbol U(t) ]
\end{align}
When we insert Eq.~(\ref{eq:app_8}) in Eq.~(\ref{eq:lyap_6}), we have
\begin{align} \label{eq:lyap_7}
& \Delta (\boldsymbol U(t))  \leq B + 2\sum_{i \in \Nset} \sum_{s \in \Sset} U_{i}^{s}(t) \lambda_{i}^{s} - 2 E[ \sum_{i \in \Nset} \sum_{s \in \Sset} U_{i}^{s}(t) \nonumber \\
& (\sum_{j \in \Nset} \oset{*}{f}_{i,j}^{s}(t) - \sum_{j \in \Nset} \oset{*}{f}_{j,i}^{s}(t)) | \boldsymbol U(t) ] + 2|\Lset| |\Sset| K F_{\max}
\end{align}
If the vector of arrival rates are interior to the stability region, there always exist $\epsilon > 0$ such that $E[\sum_{i \in \Nset} \sum_{s \in \Sset}$ $(\sum_{j \in \Nset}\oset{*}{f}_{j,i}^{s}(t) - \sum_{j \in \Nset} \oset{*}{f}_{i,j}^{s}(t)) | \boldsymbol U(t)] \leq - (\lambda_{s} 1_{[i=o(s)]} + \epsilon)$. Substituting this into Eq.~(\ref{eq:lyap_7});
\begin{align} \label{eq:lyap_8}
\Delta (\boldsymbol U(t)) & \leq B + 2|\Lset| |\Sset| K F_{\max} - 2 \sum_{i \in \Nset} \sum_{s \in \Sset} U_{i}^{s}(t) \epsilon
\end{align}
The time average of Eq.~(\ref{eq:lyap_8}) yields;
\begin{align} \label{eq:lyap_9}
\limsup_{t\rightarrow\infty} \frac{1}{t} \sum_{\tau=0}^{t-1} \Delta (\boldsymbol U(\tau)) \leq   \frac{B + 2|\Lset| |\Sset| K F_{\max}}{2\epsilon}
\end{align} which shows that the time average of $U_{i}^{s}(t)$ is bounded. Thus, TCP-aware backpressure stabilizes the network and the total average backlog is bounded.

\end{document}